\documentclass[11pt]{article} 
\usepackage{rpmacros}
\usepackage[dvipsnames,svgnames,x11names,hyperref]{xcolor}
\usepackage{amsthm}
\usepackage{xfrac}
\usepackage[T1]{fontenc}
\usepackage{enumitem}
\usepackage{fullpage}
\usepackage{xspace}%
\usepackage[russian,english]{babel}
\allowdisplaybreaks

\usepackage{thmtools}
\usepackage[colorlinks,pagebackref]{hyperref}
\hypersetup{
  linkcolor=[rgb]{0,0,0.4},
  citecolor=[rgb]{0, 0.4, 0},
  urlcolor=[rgb]{0.6, 0, 0}
}
\usepackage{mathrsfs,bbm}
\usepackage{todonotes}
\usepackage{tikz}
\usetikzlibrary{decorations.pathreplacing,backgrounds}
\usepackage{multirow}
\usepackage{setspace}
\usepackage[ruled, linesnumbered, vlined]{algorithm2e}
\let\oldnl\nl
\newcommand{\nonl}{\renewcommand{\nl}{\let\nl\oldnl}}

\usepackage[noend]{algpseudocode}
\usepackage{amsmath}
\newcommand{\algwidth}{\linewidth}
\renewcommand{\epsilon}{\varepsilon}

\usepackage{amsthm}
\usepackage{thmtools,thm-restate}

\numberwithin{equation}{section}
\declaretheoremstyle[bodyfont=\it,qed=\qedsymbol]{noproofstyle}

\declaretheorem[name=Observation,numbered=no]{observation*}

\declaretheorem[numberlike=equation]{theorem}

\declaretheorem[name=Theorem,numbered=no]{theorem*}

\declaretheorem[numberlike=equation]{lemma}
\declaretheorem[name=Lemma,numbered=no]{lemma*}

\declaretheorem[name=Corollary,numbered=no]{corollary*}

\declaretheorem[numberlike=equation]{proposition}
\declaretheorem[name=Proposition,numbered=no]{proposition*}

\declaretheorem[numberlike=equation]{claim}
\declaretheorem[name=Claim,numbered=no]{claim*}

\declaretheorem[name=Conjecture,numbered=no]{conjecture*}

\declaretheorem[name=Question,numbered=no]{question*}

\declaretheoremstyle[bodyfont=\it,qed=$\lrcorner$]{defstyle} 

\declaretheorem[numberlike=equation,style=defstyle]{definition}
\declaretheorem[unnumbered,name=Definition,style=defstyle]{definition*}

\declaretheorem[unnumbered,name=Example,style=defstyle]{example*}

\declaretheorem[unnumbered,name=Notation=defstyle]{notation*}

\declaretheorem[unnumbered,name=Construction,style=defstyle]{construction*}

\declaretheorem[numberlike=equation,style=defstyle]{remark}
\declaretheorem[unnumbered,name=Remark,style=defstyle]{remark*}

\usepackage{nth}
\usepackage{intcalc}
\usepackage{etoolbox}
\usepackage{xstring}

\usepackage{ifpdf}
\ifpdf
\else
\usepackage[quadpoints=false]{hypdvips}
\fi

\newcommand{\ECCCurl}[2]{http://eccc.hpi-web.de/report/\ifnumcomp{#1}{>}{93}{19}{20}#1/#2/}

\newcommand{\shortECCC}[2]{\texttt{\href{http://eccc.hpi-web.de/report/\ifnumcomp{#1}{>}{93}{19}{20}#1/#2/}{eccc:TR#1-#2}}}

\newcommand{\parseECCC}[1]{
\StrSubstitute{#1}{TR}{}[\tmpstring]%
\IfSubStr{\tmpstring}{/}{ 
\StrBefore{\tmpstring}{/}[\ecccyear]%
\StrBehind{\tmpstring}{/}[\ecccreport]%
}{
\StrBefore{\tmpstring}{-}[\ecccyear]%
\StrBehind{\tmpstring}{-}[\ecccreport]%
}%
\shortECCC{\ecccyear}{\ecccreport}}

\newif\ifdraft
\draftfalse
\ifdraft
\newcommand{\phnote}[1]{\todo[color=red!100!green!33,size=\footnotesize]{ph: #1}}
\newcommand{\sriprahladhuvacha}[1]{\todo[color=red!100!green!33,inline,size=\small]{ph: #1}}
\newcommand{\RPnote}[1]{\textcolor{BrickRed}{\guillemotleft RP: #1 \guillemotright}}
\newcommand{\MKnote}[1]{\textcolor{Orange}{\guillemotleft MK: #1 \guillemotright}}
\newcommand{\MSnote}[1]{\textcolor{Blue}{\guillemotleft MS: #1 \guillemotright}}

\else
\newcommand{\phnote}[1]{}
\newcommand{\sriprahladhuvacha}[1]{}

\newcommand{\RPnote}[1]{}
\newcommand{\MKnote}[1]{}
\newcommand{\MSnote}[1]{}

\fi
\newcommand{\enc}{{\mathsf{Mult}}}
\newcommand{\frs}{\mathsf{FRS}}

\newcommand{\ehref}[1]{\href{mailto:#1}{#1}}

\newcommand{\ignore}[1]{}

\DeclareMathOperator{\Char}{char}

\usepackage[nameinlink,capitalise]{cleveref}

\title{Fast list decoding of univariate multiplicity and folded Reed-Solomon codes} 
\author{
{Rohan Goyal\thanks{Chennai Mathematical Institute, Chennai, India, \ehref{rohang@cmi.ac.in}. Part of this work was done while visiting the Tata Institute of Fundamental Research, Mumbai.}}
\and 
{Prahladh Harsha\thanks{Tata Institute of Fundamental Research, Mumbai, India. \ehref{prahladh@tifr.res.in, mrinal@tifr.res.in, ashushankar98@gmail.com}.  Research supported by the Department of Atomic Energy, Government of India, under project 12-R\&D-TFR-5.01-0500. Research of the second author is partially supported by a Google India Research Award and that of the third author is supported in part by a Google Research grant and a DST-SERB grant.}}
\and
{Mrinal Kumar\samethanks}
\and
{Ashutosh Shankar\samethanks}}

\date{}

\begin{document}

\maketitle


\abstract{
  \sriprahladhuvacha{{\unexpanded{\unexpanded{TODO 
  \begin{enumerate}
    \item Add application to multivariate multiplicity codes
  \end{enumerate}}}}
  }
We show that the known list-decoding algorithms for univariate multiplicity and folded Reed-Solomon (FRS) codes  can be made to run in $\tilde{O}(n)$ time. 
Univariate multiplicity codes and FRS codes are natural variants of Reed-Solomon codes that were discovered and studied for their applications to list decoding. It is known that for every $\epsilon >0$, and rate $r \in (0,1)$, there exist explicit families of these codes that have rate $r$ and can be list decoded from a $(1-r-\epsilon)$ fraction of errors with constant list size in polynomial time (Guruswami \& Wang (\emph{IEEE Trans.\ Inform.\ Theory} 2013) and Kopparty, Ron-Zewi, Saraf \& Wootters (\emph{SIAM J. Comput.} 2023)). In this work, we present randomized algorithms that perform the above list-decoding tasks in $\tilde{O}(n)$, where $n$ is the block-length of the code.

Our algorithms have two main components. The first component builds upon the lattice-based approach of Alekhnovich (\emph{IEEE Trans. Inf. Theory} 2005), who designed a $\tilde{O}(n)$ time list-decoding algorithm for Reed-Solomon codes approaching the Johnson radius. As part of the second component, we design  $\tilde{O}(n)$ time algorithms for two natural algebraic problems: given a $(m+2)$-variate polynomial $Q(x,y_0,\dots,y_m) = \tilde{Q}(x) + \sum_{i = 0}^m Q_i(x)\cdot y_i$ the first algorithm solves order-$m$ linear differential equations of the form $Q\left(x, f(x), \frac{df}{dx}, \dots,\frac{d^m f}{dx^m}\right) \equiv 0$ while the second solves functional equations of the form $Q\left(x, f(x), f(\gamma x), \dots,f(\gamma^m x)\right) \equiv 0$, where $m$ is an arbitrary constant and $\gamma$ is a field element of sufficiently high order. These algorithms can  be viewed as generalizations of classical $\tilde{O}(n)$ time algorithms of Sieveking (\emph{Computing} 1972) and Kung (\emph{Numer.\ Math.} 1974) for computing the modular inverse of a power series, and might be of independent interest.

}
\newpage
\tableofcontents
\newpage

\section{Introduction}\label{sec:intro}
An error-correcting code (or simply a code) $\cal C$ of block length $n$ over an alphabet $\Sigma$ is a subset of the set $\Sigma^n$ such that any two distinct vectors of $\cal C$ are \emph{far} from each other in Hamming distance. The distance $\delta$ of $\cal C$ is defined as the minimum fractional distance between any two distinct vectors (henceforth, referred to as codewords) in $\cal C$ and the rate $r$ of $\cal C$ is defined as the ratio $\frac{\log |{\cal C}|}{n\log {|\Sigma|}}$ and is a measure of the \emph{redundancy} of the code. It is generally desirable to have codes $\cal C$ which \emph{large} distance and \emph{high} rate, and understanding the tradeoffs between these two quantities both information theoretically, as well as understanding what can be achieved by explicit constructions are problems of fundamental interest. Error correcting codes, though initially invented in the 1940s as a solution to the question of data storage and communication in the presence of noise have since then found numerous connections and applications to theory and practice alike. Both directly, and indirectly, ideas in this area continue to find relevance in complexity theory, pseudorandomness, combinatorics, cryptography and are central objects of study across these areas. 
\paragraph*{Decoding and list-decoding }
If the distance of a code $\cal C$ is $\delta$, then it follows that for any vector $v \in \Sigma^n$, there is at most one codeword $c$ of $\cal C$ at a fractional Hamming distance less than $\delta/2$ from $\cal C$, and given $v$ as input, the task of computing $c$ is referred to as (unique) decoding of $\cal C$. Designing an efficient decoding algorithm for this task for a code $\cal C$ of interest is a fundamental problem of interest. If we consider a Hamming ball of radius greater than $\delta/2$ centered at a vector $v \in \Sigma^n$, there might now be multiple codewords of $\cal C$ present in this ball, and for sufficiently large radii, this number may become  exponentially large. However, a classical result in coding theory asserts that this number continues to be polynomially bounded in $n$ as long as the radius of the Hamming ball is less than the so-called \emph{Johnson radius} of the code. The algorithmic task of computing all the codewords within such a Hamming ball is referred to as list decoding. In general, if the number of codewords of a code $\cal C$ in a Hamming ball of radius $r$ around any vector $v \in \Sigma^n$ is at most $L$, then $\cal C$ is said to be $(r, L)$-\emph{list-decodable}. Seeking efficient algorithms for list decoding well-studied families of codes is again a fundamental line of enquiry in this area of research, and the focus of this paper. 

\paragraph*{List-decodable and algebraic codes }Algebraic techniques and constructions have traditionally been of immense significance in the study of error correcting codes, with families of codes like Reed-Solomon codes and Reed-Muller codes being perhaps one of the most well-studied families of codes over the years. For Reed-Solomon codes, the alphabet $\Sigma$ is a finite field $\F$ and the codewords correspond to the evaluations of all univariate polynomials of degree $d$ over $\F$ (or a subset of size $n$ of $\F$). These codes can be seen to have fractional distance $\delta = 1 - d/n$ and rate $r = (d+1)/n$ and are thus optimal in the sense that they achieve the so-called \emph{Komamiya-Joshi-Singleton bound} (popularly referenced to as the \emph{Singleton bound}). Reed-Muller codes are the multivariate analogues of Reed-Solomon codes where codewords correspond to the evaluations of multivariate polynomials of low degree over inputs from the underlying field. Some of these algebraic codes are at the heart of the notions of \emph{local decoding}, \emph{local correction} and \emph{local testing} that in addition to being fundamental notions on their own, have also found numerous applications in complexity theory and pseudorandomness. 

Yet another example of this fruitful alliance of algebraic techniques and coding theory is in the study of list decoding of codes. In a beautiful and influential work in 1996, Sudan \cite{Sudan1997} showed that Reed-Solomon codes can be efficiently list-decoded when the fraction of errors is less than $1 - \sqrt{2d/n}$, which is far beyond the unique decoding regime of $(1-d/n)/2$ fraction errors when $d = \alpha n$ for small enough constant $\alpha < 1$. Guruswami and Sudan \cite{GuruswamiS1999} subsequently improved this to efficient list decoding with less than $1 - \sqrt{d/n}$ fraction errors which matches the Johnson radius for these codes. The combinatorial and algorithmic list decodability of Reed-Solomon codes beyond the Johnson radius is far from understood and continues to be a problem of great interest, including in the light of some very recent work (e.g., see \cite{ShangguanT2023,AlrabiahGL2023} and references therein). 

Owing to their central role in coding theory and computer science, the question of designing faster (ideally nearly-linear-time) decoding algorithms for Reed-Solomon codes has also received considerable attention. In the unique decoding regime, a nearly-linear-time decoder for Reed-Solomon codes was designed essentially by showing that the classical Berlekamp-Massey algorithm \cite{Berlekamp1967, Massey1969} for decoding these codes has a nearly-linear-time implementation (see \cite{ReedSTW1978} and references therein).  In the list-decoding regime, a similar result was shown in a work of Alekhnovich \cite{Alekhnovich2005} that gave an algorithm for decoding Reed-Solomon codes from $1 - \sqrt{d/n} - \epsilon$ fraction errors in nearly-linear time, for every constant $\epsilon > 0$. Alekhnovich's overall approach and techniques that were based on the study of appropriate lattices \footnote{See \autoref{defn:lattices} for a formal definition of a lattice over a univariate polynomial ring.} over the univariate polynomial ring are crucial to our work as well. We discuss this in greater detail in our proof overview (\autoref{sec:proof overview}). 

In subsequent work towards understanding list decodability beyond the Johnson radius, natural variants of Reed-Solomon codes, such as the Folded Reed-Solomon (FRS) codes and Multiplicity codes were studied. In a seminal work, Guruswami \& Rudra \cite{GuruswamiR2008}, building on the work of Parvaresh and Vardy \cite{ParvareshV2005}, showed that FRS codes are (algorithmically) list-decodable \emph{far} beyond the Johnson radius, in fact all the way upto capacity. In subsequent works, Guruswami \& Wang \cite{GuruswamiW2013} and Kopparty \cite{Kopparty2015} showed that multiplicity codes also have this amazing list-decoding property upto capacity. More recently, Kopparty, Ron-Zewi, Saraf \& Wootters \cite{KoppartyRSW2023} showed the surprising fact that these codes are list-decodable with \emph{constant} list size even as they approach list-decoding capacity. Moreover, they showed that the list-decoding algorithm of Guruswami \& Wang can be combined with an efficient \emph{pruning} procedure that gives an efficient list-decoding algorithm for FRS and multiplicity codes up to capacity with constant list size. In this work, we design significantly faster algorithms for list decoding these codes from close to list-decoding capacity. 

We start by defining FRS and multiplicity codes, before stating our results formally.

\subsection{Folded Reed-Solomon and univariate multiplicity codes}

FRS codes are a variant of Reed-Solomon codes, first defined by Krachkovsky \cite{Krachkovsky2003} and are the first codes that were shown to be efficiently list-decodable upto capacity by Guruswami and Rudra \cite{GuruswamiR2008}.
\begin{definition}[Folded Reed-Solomon \cite{Krachkovsky2003,GuruswamiR2008}]\label{def:frs-codes}
Let $\F$ be a finite field of size $q$ and let $n, d, s$ be positive integers satsifying $d/s < n < (q-1)/s$. Let $\gamma \in \F$ be a primitive element of $\F$ and let $\alpha_1, \alpha_2, \ldots, \alpha_n$ be distinct elements from the set $\{\gamma^{si} \colon i \in \{0, 1, \ldots, ((q-1)/s) - 1\}\}$. Then, an FRS code with folding parameter $s$ for degree $d$ polynomials is defined as follows. 

The alphabet of the code is $E = \F^{s}$ and the block length is $n$, where the coordinates of a codeword are indexed by $1, 2, \ldots, n$. The message space is the set of all univariate polynomials of degree at most $d$ in $\F[x]$. And, the encoding of such a message $f \in \F[x]$, denoted by $\frs_{s}(f) \in E^n$ is defined as 
\[
\frs_{s}(f)|_{i} := \left(f(\alpha_i), f(\gamma\alpha_i), \ldots, f(\gamma^{s-1}\alpha_i)\right) \, \qedhere.
\]
\end{definition}

Univariate multiplicity codes are yet another variant of Reed-Solomon codes and were defined by Rosenbloom \& Tsfasman \cite{RosenbloomT1997} and Nielson \cite{Nielsen2001} and were later generalized to the multivariate setting by Kopparty, Saraf and Yekhanin \cite{KoppartySY2014} who studied them from the perspective of local decoding. In this paper, we only work with univariate multiplicity codes that we now formally define. 
\begin{definition}[Multiplicity codes \cite{RosenbloomT1997,Nielsen2001,KoppartySY2014}]\label{def:multiplicity codes}
Let $n, d, s$ be positive integers satisfying $ds < n$, $\F$ be a field of characteristic zero or at least $d$ and size at least $n$, and let $\alpha_1, \alpha_2, \ldots, \alpha_n$ be distinct elements of $\F$. Then, univariate multiplicity codes with multiplicity parameter $s$ for degree $d$ polynomials is defined as follows. 

The alphabet of the code is $E = \F^{s}$ and the block length is $n$, where the coordinates of a codeword are indexed by $1, 2, \ldots, n$. The message space is the set of all univariate polynomials of degree at most $d$ in $\F[x]$. And, the encoding of such a message $f \in \F[x]$, denoted by $\enc_{s}(f) \in E^n$ is defined as 
\[
\enc_{s}(f)|_{i} := \left(f(\alpha_i), f^{(1)}(\alpha_i), \ldots, f^{(s-1)}(\alpha_i)\right) \, ,
\]
where $f^{(j)}(\alpha_i)$ is the evaluation of the $j^{th}$-order derivative\footnote{\label{fn:mult-ideal-defn}This definition in terms of (standard) derivatives requires the characteristic of the field to be 0 or at least $d$. All known capacity-achieving list-decoding results for univariate multiplicity codes require large characteristic (previous works \cite{Kopparty2015,GuruswamiW2013,KoppartyRSW2023} as well as our work). For this reason as well as ease of presentation, we work with standard derivatives. A more general definition, which works for all characteristics, can be given in terms of Hasse derivaties or using the ideal-theoretic framework as follows: the alphabet $E = \F^s$ is interpreted as $\F_{<s}[x]$ (polynomials of degree strictly less than $s$) and the $i^{th}$ symbol of the encoding $\enc_{s}(f) \in E^n$ of message $f \in \F[x]$, is defined as 
$\enc_{s}(f)|_{i} := f(x) \mod (x-\alpha_i)^s$.} of $f$ on the input $\alpha_i$. 
\end{definition}

Both these families of codes have rate $d/sn$, alphabet size $|\F|^s$ and distance $1-d/sn$, and are known to have some remarkable list-decoding properties. In particular, for every $\epsilon > 0$ and for sufficiently large $s = s(\epsilon)$, these are both known to be efficiently list-decodable from $1-d/sn-\epsilon$ fraction of errors. For FRS codes, this was shown by Guruswami \& Rudra \cite{GuruswamiR2008} (also a different analysis by Guruswami \& Wang \cite{GuruswamiW2013}) and for multiplicity codes, this was shown independently by Guruswami \& Wang \cite{GuruswamiW2013} and by Kopparty \cite{Kopparty2015}. A recent work of Bhandari, Harsha, Kumar \& Sudan \cite{BhandariHKS2024-affineFRS} shows that the framework of Guruswami \& Wang \cite{GuruswamiW2013} can be abstracted out into a unified algorithm for decoding multiplicity and FRS codes, as well as some of the other variants, e.g., additive or affine FRS codes. 

Kopparty, in his survey on multiplicity codes \cite[Section 7, Item (3)]{Kopparty2014} asks the open question if univariate multiplicity codes can be list decoded in nearly-linear time up to the Johnson radius? In particular, can one emulate the improvements that Alekhnovich's algorithm \cite{Alekhnovich2005} obtains for the Guruswami-Sudan list-decoding algorithm for Reed-Solomon codes to the setting of univariate multiplicity codes. A recent work of Kopparty, Saraf, Ron-Zewi and Wootters \cite{KoppartyRSW2023} showed that both FRS and univariate multiplicity codes approach list-decoding capacity while maintaining a constant (depending upon the parameter $\epsilon$ where $1-d/sn -\epsilon$ is the fraction of errors) list size. Given such linear (in fact constant) sized bounds on the list size, one may ask if the following strengthening of Kopparty's question also holds true: can univariate multiplicity and FRS codes be list decoded in nearly-linear time all the way up to capacity? The main contribution of our work is a positive resolution of these open questions. 

\subsection{Our results}\label{sec:results}
\paragraph*{Results on list decoding} Our main results are summarized in the following theorems. We start with the statement for multiplicity codes.
\begin{theorem}\label{thm:intro-main-large-multiplicity}
For every $\epsilon > 0$, there exists a natural number $s_0$ such that for all $s > s_0$, degree parameter $d$, block length $n$, field $\F$ of characteristic zero or larger than $d$, the following is true. 

There is a randomized algorithm that runs in time $O\left(n\cdot \poly(s\log n \log |\F| (sn/d)^{\poly(1/\epsilon)})\right)$ and with high probability, list decodes univariate multiplicity code with multiplicity parameter $s$ for degree $d$ polynomials over $\F$ from $\left(1-\frac{d}{sn} - \epsilon\right)$ fraction of errors. 
\end{theorem}
We now state the analogous result for FRS codes. 
\begin{theorem}\label{thm:intro-frs-main-large-folding}
For every $\epsilon > 0$, there exists a natural number $s_0$ such that for all $s > s_0$, degree parameter $d$, block length $n$, field $\F$ such that $d/s < n < (|\F|-1)/s$, the following is true. 

There is a randomized algorithm that runs in time $O\left(n\cdot \poly(s\log n \log |\F| (sn/d)^{\poly(1/\epsilon)})\right)$ and with high probability, list decodes Folded Reed-Solomon codes with folding parameter $s$ for degree $d$ polynomials over $\F$ from $\left(1-\frac{d}{sn} - \epsilon\right)$ fraction errors. 
\end{theorem}

Both the above results require the parameter $s$ to be large compared to $\epsilon$. This constraint is essentially due to the current state of our understanding of list decoding for these codes. In particular, we don't know if these codes are list-decodable up to capacity for small $s$. As a warm-up for some of the ideas in our proofs, we show that for all $s$ these codes can be decoded up to a distance $\epsilon$ of the so-called Johnson radius in nearly-linear time. This algorithm is 
a 
generalization of a result of Alekhnovich \cite{Alekhnovich2005} for decoding Reed-Solomon codes up to a distance $\epsilon$ of the Johnson radius. 
\begin{theorem}\label{thm:intro-main-all-multiplicity}
For every $\epsilon > 0$, degree parameter $d$, block length $n$, multiplicity parameter $s$, and field $\F$ of any characteristic, the following is true. 

There is a deterministic algorithm that runs in time $(n\cdot \poly(s \log n \log |\F| (sn/d)))$, and list decodes order-$s$, univariate multiplicity code for degree $d$ polynomials over $\F$ from $\left(1-(d/sn)^{1/2} - \epsilon\right)$ fraction errors. 
\end{theorem}
A similar result holds for FRS codes and we skip the formal statement here. 

Since $s$ and $\epsilon$ are constants, the algorithms in \autoref{thm:intro-main-large-multiplicity}, \autoref{thm:intro-frs-main-large-folding} and \autoref{thm:intro-main-all-multiplicity} run in nearly-linear time in the block length $n$ provided that the rate $d/sn$ of the code is a constant. 
\paragraph*{Fast solvers for differential and functional equations }
On the way to our results, we design nearly-linear-time algorithms for solving certain linear differential and functional equations that arise naturally in the context of decoding multiplicity and FRS codes. These algorithms build upon the techniques like Fast Fourier transform and Newton iteration from computational algebra and to our knowledge do not appear to be known in literature. We suspect that these results might be of independent interest and might have applications besides their current application to list decoding. We state the main theorems starting with the result on differential equations. 
 
\begin{theorem}\label{thm:intro-fast-differential-equation-solver}
Let $\F$ be a finite field of characteristic greater than $d$ or zero, and let \[Q(x, y_0, \ldots, y_m) = \tilde{Q}(x) + \sum_{i = 0}^m Q_i(x)y_i\] be a non-zero polynomial with $x$-degree at most $D$. Then, the set of polynomials $f(x) \in \F[x]$ of degree at most $d$ that satisfy 
\[
Q\left(x, f, f^{(1)} \ldots, f^{(m)}\right) \equiv 0
\]
form an affine space dimension at most $m$. Furthermore, there is a deterministic algorithm that when given $Q$ as an input via its coefficient vector, and the parameter $d$, performs at most $\tilde{O}((D+d)\poly(m))$ field operations and outputs a basis for this affine space. 
\end{theorem}  
We now state an analogous result for solving certain functional equations.
\begin{theorem}\label{thm:frs-fast-functional-equation-solver-original-intro}
Let $\F$ be any field, let $\gamma \in \F$ be an element of order greater than $d$, and let \[Q(x, y_0, \ldots, y_m) = \tilde{Q}(x) + \sum_{i = 0}^m Q_i(x)y_i\] be a non-zero polynomial with $x$-degree at most $D$. Then, the set of polynomials $f(x) \in \F[x]$ of degree at most $d$ that satisfy 
\[
Q\left(x, f(x), f(\gamma x), \ldots, f(\gamma^{m} x)\right) \equiv 0
\]
form an affine space of dimension at most $m$. Furthermore, there is a deterministic algorithm that when given $Q$ as an input via its coefficient vector, and the parameter $d$, performs at most $\tilde{O}((D+d)\poly(m))$ field operations and outputs a basis for this affine space. 
\end{theorem}  

Earlier algorithms for solving linear differential equations exist for settings different from ours or a running time that exceeds our requirements \cite{AbramovK1991, Singer1991}. These mostly proceed via finding intermediate power series or rational solutions. Bostan, Cluzeau and Salvy \cite{BostanCS2005} give a nearly-linear-time algorithm but in the setting where the degree of the coefficients is small and the degree of the solution is growing (in our case, the degree of the coefficients depends on the block length of the code).\phnote{Reviewer's comment: \emph{Explain what you mean by “degree of the coefficients” and “degree of the solution”.}}

\subsection{Related Work}

In this section, we review related work in the two different contexts: (a) nearly-linear-time decoding and (b) list-size bounds.

\paragraph*{Nearly-linear-time decoding algorithms:} There is a rich literature in coding theory trying to obtain nearly-linear-time decoding algorithms. As mentioned earlier, the classical Berlekamp-Massey algorithm \cite{Berlekamp1967,Massey1969} for unique-decoding Reed-Solomon codes can be implemented in nearly-linear time \cite{ReedSTW1978}. Alekhnovich \cite{Alekhnovich2005} obtained a nearly-linear-time implementation of the Guruswami-Sudan algorithm \cite{GuruswamiS1999} for list decoding Reed-Solomon codes up to the Johnson radius. Subsequent to the discovery that FRS and univariate multiplicity codes are list-decodable up to capacity \cite{GuruswamiR2008,Kopparty2015,GuruswamiW2013}, Hemenway, Ron-Zewi \& Wootters \cite{HemenwayRW2020} constructed codes that were list-decodable all the way up to capacity in \emph{probabilistic} nearly-linear time, more specifically $n^{1+o(1)}$ time where $n$ is the blocklength of the code. In a followup work, Kopparty, Resch, Ron-Zewi, Saraf \& Silas \cite{KoppartyRRSS2021} proved a similar result for codes that were list-decodable all the way up to capacity in \emph{deterministic} nearly-linear time. Hemenway, Ron-Zewi \& Wootters \cite{HemenwayRW2020} and Kopparty, Resch, Ron-Zewi, Saraf \& Silas \cite{KoppartyRRSS2021} obtain such codes by a clever application of expander distance amplification technique of Alon, Edmonds and Luby \cite{AlonEL1995,AlonL1996} on high-rate list-decodable codes. We remark that these results which construct the first nearly-linear-time list-decodable codes all the way upto capacity do so by constructing tailor-made codes (starting from FRS/Multiplicity like codes of very small block-length) while our nearly-linear-time list-decoding algorithms are for the standard FRS and univariate multiplicity codes. Furthermore, our algorithms run in time $O(n \cdot \poly\log n)$ while to the best of our knowledge, the algorithms of \cite{HemenwayRW2020,KoppartyRRSS2021} cannot be made to run in time $O(n \cdot \poly \log n)$. That said, while our nearly-linear-time algorithms are randomized, the nearly-linear-time algorithms of Kopparty, Resch, Ron-Zewi, Saraf \& Silas \cite{KoppartyRRSS2021} are deterministic.

Sipser and Spielman constructed expander codes, which were the first family of codes that are unique decodable in linear time \cite{SipserS1996}. Guruswami and Indyk \cite{GuruswamiI2003} used expander-based constructions to construct the first family of codes that were list-decodable in linear time. It would be be interesting if similar expander-based constructions can be used to construct codes that are list-decodable all the way up to the Johnson radius or even capacity in linear time. 

\paragraph{Bounds on list size:} The initial works on list-decodable codes up to capacity \cite{GuruswamiR2008,Kopparty2015,GuruswamiW2013} obtained only polynomial-sized list bounds. This was dramatically improved by Kopparty, Ron-Zewi, Saraf and Wootters \cite{KoppartyRSW2023} who showed that FRS and univariate multiplicity codes are list-decodable up to capacity with constant-sized lists. There has been a lot of work towards improving the list size of list-decodable codes \cite{KoppartyRSW2023,KoppartyRRSS2021,Tamo2024}. In this work, we do not focus on list-size bounds. The list-size bounds we obtain are inherited from the results of Kopparty, Ron-Zewi, Saraf and Wootters \cite{KoppartyRSW2023} (and subsequent improvements due to Tamo \cite{Tamo2024}). We refer the interested reader to previous works (\cite{KoppartyRSW2018,KoppartyRRSS2021,Tamo2024}) for a survey on list-size bounds.

\section{Overview of the proofs}\label{sec:proof overview}
In this section, we give a short overview of the proofs of our results. Our algorithms and analysis for multiplicity codes and FRS codes are very similar to each other, and therefore, in this overview we focus only on multiplicity codes. Even later in this paper, we discuss the proofs for multiplicity codes in detail, whereas we sketch out the arguments for FRS codes. Moreover, we focus on the case of capacity-achieving decoding, i.e.,\autoref{thm:intro-main-large-multiplicity} and \autoref{thm:intro-frs-main-large-folding}.  

Our decoding algorithm for multiplicity codes is essentially a faster implementation of the list-decoding algorithm of Guruswami \& Wang \cite{GuruswamiW2013}. We follow the outline of their algorithm, observe the computational bottlenecks, and design faster alternatives for those steps. We start this overview with a brief outline of Guruswami \& Wang's algorithm. 

Let $R = (\alpha_i, \beta_{i,0}, \ldots, \beta_{i, s-1})_{i = 1}^n$ be the received word 
that we wish to list decode from. The algorithm in \cite{GuruswamiW2013} essentially has two main steps, which are the first two steps mentioned below. The third step below is from a subsequent work of Kopparty, Ron-Zewi, Saraf and Wootters \cite{KoppartyRSW2023} (and a follow up work of Tamo \cite{Tamo2024}) who showed how to reduce the list size obtained in \cite{GuruswamiW2013} to a constant. 

\subsection[Outline of the known algorithms]{Outline of the known algorithms \cite{GuruswamiW2013,KoppartyRSW2023}}
\paragraph*{Constructing a differential equation : } The aim in this step is to find an $(m+2)$-variate non-zero polynomial $Q$ of the form $Q(x, \vecy) = \tilde{Q}(x) + \sum_{i=0}^m Q_i(x) y_i$, where each $Q_i$ has low degree (denoted by $D$) and that $Q$ satisfies some interpolation constraints. The parameter $m$ is chosen carefully depending on $s$ and $\epsilon$ and for this discussion, can just be thought of something smaller than $s$. The interpolation constraints are imposed in a way that ensures that for any degree $d$ univariate $f(x)$ whose multiplicity code encoding is \emph{close enough} to the received word $R$, $f$ satisfies the differential equation \[Q(x, \partial f) := \tilde{Q}(x) + \sum_i Q_i(x) \cdot f^{(i)}(x) \equiv 0 \, .\] To this end, for every $j \in \{1, \ldots, n\}$, we impose constraints of the form 
\begin{align*}
\tilde{Q}(\alpha_j) + \sum_i Q_i(\alpha_j)\cdot \beta_{j, i} &= 0 \, , \text{and} \\
\tilde{Q}^{(1)}(\alpha_j) + \sum_i \left(Q_i^{(1)}(\alpha_j)\cdot \beta_{j,i} + Q_i(\alpha_j)\cdot \beta_{j+1,i}\right) &= 0\,.
\end{align*}
The constraints above have been engineered in a way that ensures that if the encoding of $f$ and $R$ agree at a point $\alpha_i$, then $Q(x, \partial f)$ vanishes at $\alpha_i$ and the first derivative of $Q(x, \partial f)$ vanishes at $\alpha_i$. More generally, about $s-m$ similar such constraints are imposed for every $i$ that ensure that every point $\alpha_i$ of agreement,  $Q(x, \partial f)$ vanishes with multiplicity at least $s-m$. Thus, if the number of points of agreements is sufficiently large as a function of $D, d$, we get that $Q(x, \partial f)$ must be identically zero as a polynomial. In particular, in order to solve the list-decoding question, it suffices to solve the differential equation $Q(x, \partial f) \equiv 0$ for solutions of degree at most $d$. 

\paragraph*{Solving the differential equation $Q(x, \partial f) \equiv 0$ : } Guruswami \& Wang then solve this differential equation, and show that there are at most polynomially many degree-$d$ solutions to it, and that the list of solutions can be computed in polynomial time. To this end, their main observation is that the space of solutions is an affine space of dimension at most $m$, and it suffices to compute a basis of this subspace. To compute the basis, they open up the structure of this linear system, and notice that the system is essentially lower triangular with at most $m$ zeroes on the diagonal. 

\paragraph*{Pruning to a constant-sized list:} In their work \cite{KoppartyRSW2023}, Kopparty, Saraf, Ron-Zewi and Wootters improved the bound on the size of the list output by Guruswami \& Wang to a constant. They do this by essentially showing that for a code like multiplicity code or an FRS code that has good distance, a constant dimensional affine subspace can have at most a constant number of codewords that are \emph{close enough} to a given received word. In addition to this, they also gave a simple randomized algorithm to obtain this pruned list. 

\subsection{A faster algorithm}

For our algorithm, we present alternative algorithms that implement the first two steps above in nearly-linear time. Naively, the first step above involves solving a linear system with at least $n$ constraints, and the best state of art general purpose linear system solvers can do this is in roughly matrix multiplication time (so at least $\Omega(n^2)$ time). Similarly it is \textit{a priori} unclear if the step on solving differential equations can be done in better than $O(n^2)$ time. The final pruning step of Kopparty, Ron-Zewi, Saraf and Wootters \cite{KoppartyRSW2023} and Tamo \cite{Tamo2024} happened to be nearly-linear time as it is, and we use this step off-the-shelf from their work without any changes. We note that the randomness in our final algorithm only comes from the randomness in the algorithm of Kopparty, Ron-Zewi, Saraf and Wootters \cite{KoppartyRSW2023}. As an intermediate statement, we get the following theorem, which when combined with the aforementioned pruning procedure gives us \autoref{thm:intro-main-large-multiplicity}.

\begin{theorem}\label{thm:main-large-multiplicity}
For every $\epsilon > 0$, there exists a natural number $s_0$ such that for all $s > s_0$, degree parameter $d$, block length $n$, field $\F$ of characteristic zero or larger than $d$, the following is true. 

There is a deterministic algorithm that takes as input a received word $R$ and after at most  $O\left(n\cdot \poly(s\log n\poly(1/\epsilon))\right)$ field operations outputs a basis for an affine space $A$ of degree $d$ polynomials of dimension $O(1/\epsilon)$ such that if $\enc_{s}(f)$ has fractional Hamming distance at most $\left(1-\frac{d}{sn} + \epsilon\right)$ from $R$ for a polynomial $f$ of degree at most $d$, then $f$ is contained in $A$.

\end{theorem}
We now briefly discuss the ideas in our faster algorithms for constructing the differential equation and for solving them. 

\paragraph*{Fast construction of the differential equation: }To construct the interpolating polynomial $Q(x, \vecy) = \tilde{Q}(x) + \sum_{i} Q_i(x)y_i$ satisfying the linear constraints described above in nearly-linear time, we rely on the ideas of Alekhnovich \cite{Alekhnovich2005}. In particular, we define an appropriate lattice over the polynomial ring $\F[x]$ with the properties that any non-zero polynomial in this lattice satisfies the linear constraints of interest, and moreover that the shortest vector in the lattice gives us a polynomial $Q$ of low enough degree for our applications. At this point, we invoke an algorithm of Alekhnovich to exactly compute the shortest vector of this lattice, which gives us the desired $Q$. We remark on the two salient points of departure from Alekhnovich's algorithm and analysis. First, the definition of the lattice that is suited for decoding multiplicity codes (both for decoding upto Johnson radius as well as upto capacity) is different from that of Alekhnovich. Second, Alekhnovich proves that there exists a short vector in the lattice by demonstrating that the lattice constructed by him is identical to the one constructed by Guruswami and Sudan which is known to have short vectors. We instead follow a more direct approach; we demonstrate the existence of a short vector by a direct application of Minkowski's theorem. 
The details can be found in \autoref{sec:mult-differential-eqn-const}.

\paragraph*{Fast solver for linear differential equations: }
We now outline the main ideas in our proof of \autoref{thm:intro-fast-differential-equation-solver}. For ease of notation, we confine ourselves to solving an equation of the form $Q(x,\partial f) = \tilde{Q}(x) + Q_0(x)f(x) + Q_1(x)f^{(1)}(x) \equiv 0$, i.e.,the parameter $m = 1$. If the polynomial $Q_1$ is identically zero, then, we are solving $\tilde{Q} + Q_0f \equiv 0$. We note that in this case, $Q_0(x)$ can be assumed to have a non-zero constant term, essentially without loss of generality. To see this, observe that for any $i$, if $x^i$ divides $Q_0(x)$, then, it must also divide $\tilde{Q}(x)$ or else the equation has no solution. Thus, we can simply divide the whole equation $\tilde{Q} + Q_0f \equiv 0$ by the largest power of $x$ that divides $Q_0$, and it can be assumed to have a non-zero constant term. In this case, the solution $f$ is just equal to $f = -\tilde{Q}/Q_0$. Thus, to compute this, it suffices to compute the inverse of $Q_0$ in the ring $\F[[x]]$ of power series. Moreover, since we are interested in solutions $f$ of degree at most $d$, it suffices to compute the inverse modulo $x^{d+1}$. There are classical nearly-linear-time algorithms known for this problem, e.g., see  Sieveking \cite{Sieveking1972} and Kung \cite{Kung1974} and over the years, these algorithms have found many applications. These algorithms are based on ideas like Fast Fourier transform and Newton iteration. In fact in his work on fast list decoding of Reed-Solomon codes, Alekhnovich uses an algorithm of Roth \& Ruckenstein \cite{RothR2000} based on very similar ideas to solve the polynomial equations that arise there. 

We would now like to emulate the above approach to larger $m$, i.e., $m > 1$. However, it is unclear to us if the ideas there immediately extend to the case of differential equations that arise here. The fact that we no longer have unique solutions to the differential equations causes a problem when one implements the above divide-and-conquer approach in a naive manner for the following reason. Suppose at each step of the divide-and-conquer the number of solutions doubles, then the total number of solutions would be prohibitively large if one unravels the recursion. To get around this issue, we show that given any differential equation $Q$, there is a related differential equation $Q^\dagger$ which has a unique solution and from whose unique solution we can construct all the solutions of $Q$. Below, we give a high-level overview of this approach.

Going forward, we assume that the polynomial $Q_1$ is non-zero. Additionally, up to a translation of the $x$ variable, we also assume that $Q_1$ has a non-zero constant term. In this case, Guruswami and Wang had shown that the affine space of solutions has dimension $1$ and if we fix the constant term of any solution, then there is a unique polynomial $f$ of degree $d$ with this specified constant term that is a solution. Let us fix this constant term to be $1$ and look for solutions of the form $f = 1 + xg$ where $g$ is a polynomial of degree less than $d$. Thus, it suffices to solve for $g$.  Using the properties of the derivative operator, we get
\begin{align*}
Q(x, f, f^{(1)}) &= \tilde{Q} + Q_0\cdot f + Q_1\cdot f^{(1)} \\
&= \tilde{Q} + Q_0 \cdot(1 + xg) + Q_1\cdot (g + xg^{(1)})\\
&=\left(\tilde{Q} + Q_0\right) + (x\cdot Q_0 + Q_1)\cdot g + x\cdot Q_1\cdot g^{(1)}\,.
\end{align*}
Thus, if $f$ satisfies $Q(x, \partial f) \equiv 0$, it must be the case that $g$ satisfies $Q^{\dagger}(x, \partial g) \equiv 0$ for  \[Q^{\dagger}(x, y_0, y_1) := \left(\tilde{Q} + Q_0\right) + (x\cdot Q_0 + Q_1)\cdot y_0 + x\cdot Q_1\cdot y_1  \, .\]
While $Q^{\dagger}$ looks similar to $Q$ in overall form, there are small changes in structure when compared to $Q$: for instance, the coefficient $x\cdot Q_1$ of $y_1$ has zero as its constant term. It turns out that these seemingly small changes in overall structure make $Q^{\dagger}$ much nicer to work with. In particular, we observe that equations of this form now have unique solutions. Moreover, the uniqueness of solution and the form of $Q^{\dagger}$ makes it very amenable to a divide-and-conquer-style algorithm based on Newton iteration. We show that in order to solve $Q^{\dagger}(x, \partial g) \equiv 0$ for a $g$ of degree less than $d$, it suffices (up to some nearly-linear computational overhead) to solve two equations of a similar form, where we are now  looking for solutions of degree $d/2$. This observation then immediately gives a recursive algorithm, where the recursion is on the degree of solution of interest. One slightly technical point to note is that for this recursion to work, we end up working with the modular equation $Q^{\dagger}(x, \partial g) \equiv 0 \mod x^k$ for sufficiently large $k$, which is also the case for many of the other applications of Newton iteration-like techniques. While the outline of the above is quite clean, the details end up being slightly technical and we describe the detailed algorithm and its analysis in \autoref{sec:solving-differential-equations}. 

We also note that attempting a similar strategy while directly working with $Q$ and not $Q^{\dagger}$ seems like a natural thing to try. However, note that the differential equation for $Q$ (unlike that for $Q^{\dagger}$) does not have a unique solution and hence a recursive approach to solve the differential equation for $Q$ directly requires a lot of book-keeping and other issues to ensure that the number of solutions do not blow up. Working with $Q^{\dagger}$ gets around these problems and is crucial for our proof. 

\subsection*{Organization of the paper}
The rest of the paper is organized as follows. 

We start with some notations and preliminaries in \autoref{sec:prelims}. In \autoref{sec:mult-differential-eqn-const}, we discuss a  fast algorithm for construction of differential equations for decoding multiplicity codes. Our algorithm for solving these linear differential equations is discussed in \autoref{sec:solving-differential-equations}. We combine these results to conclude the proof of \autoref{thm:intro-main-large-multiplicity} in \autoref{sec:mult-list-decoding}.  We also discuss a natural extension of Alekhnovich's algorithm in \autoref{sec:johnson} which gives us a proof of \autoref{thm:intro-main-all-multiplicity}.

Finally, in \autoref{sec:frs-constructing-functional-equations} and \autoref{sec:frs-solving-functional-equations}, we discuss some of the main ideas needed for the proof of \autoref{thm:intro-frs-main-large-folding}. Since the ideas are quite similar to those in the proof of \autoref{thm:intro-main-large-multiplicity}, we focus on the differences, and some of the proofs in these sections are left as sketches. 
\section{Notations and Preliminaries}\label{sec:prelims}

\subsection{Notations}
We summarise some of the notations that we use.
\begin{itemize}
\item We use $\F$ to denote a field. Throughout this paper, $\F$ is finite, unless otherwise stated.
\item We use boldface letters like $\vecy$ to denote a tuple of vectors. The arity of the tuple is generally clear from the context. 
\item Given a set of vectors $\{v_0, v_1, v_2, \ldots, v_k\} \subseteq \F^m$, their affine span is defined as the set $\{\sum_{i=0}^k \alpha_iv_i : \alpha_i \in \F, \sum_{i=0}^k\alpha_i = 1\}$. 
\item For an affine space $A$ in $\F^m$, where $\F$ is the underlying field, we define the canonical linear space corresponding to $A$ to be the vector space $\tilde{A}$ over $F$ such that there is a vector $v \in \F^m$ with $A = v + \tilde{A}$. The dimension of $A$ is defined to be the dimension of $\tilde{A}$.
\item For every affine space $A \subset \F^m$ of dimension $k$, there is a set $\{v_0, v_1, v_2, \ldots, v_k\}$ of $k+1$ vectors in $A$ such that $\{v_1-v_0,\ldots v_k-v_0\}$ are linearly independent over the underlying field $\F$ such that $A$ equals the affine span of $\{v_0, v_1, v_2, \ldots, v_k\}$. We refer to any such set of vectors as a basis of $A$. 
\item For a polynomial $Q(x, y)$ in $\F[x, y]$, the $(a, b)$-weighted degree of $Q$ is defined as $a \deg_x(Q) + b \deg_y(Q)$.
\end{itemize}



\subsection{Fast polynomial arithmetic}
\begin{lemma}[Taylor expansion of polynomials]\label{lem:taylor-for-polynomials}
Let $\F$ be a field and $d$ be a natural number such that the characteristic of $\F$ is either zero or larger than $d$. Then, for every univariate polynomial $f(x) \in \F[x]$ of degree $d$, we have that
\[
f(x+z) = f(x) + f^{(1)}(x)z + \frac{f^{(2)}(x)z^2}{2!} + \cdots + \frac{f^{(d)}(x)z^d}{d!} \, .
\]
\end{lemma}
The following simple lemma will be useful for our proofs. 
\begin{lemma}\label{lem:computing-derivatives-fast}
Let $f(x)$ be a univariate polynomial over a field $\F$ of degree at most $d$ and let $m \leq d$ be a natural number. Then, there is an algorithm that requires at most $O(dm)$ field operations and outputs the coefficient vectors of the derivatives $f^{(1)}, \ldots, f^{(m)}$ of $f$, when given the coefficient vector of $f$ as input.  
\end{lemma}
\begin{proof}
We can compute these derivatives by computing the first $m$ derivatives of each of the monomials $1, x, \ldots, x^d$. Now, to compute the first $m$ derivatives of a monomial $x^k$ requires us to compute the field elements $k, k(k-1), k(k-1)(k-2), \ldots, k(k-1)(k-2)\cdots (k-m+1)$, which requires at most $O(m)$ field operations. Thus, all the first $m$ derivatives of $f$ can be computed using at most $O(dm)$ field operations.
\end{proof}

\subsection[Computing shortest vector in polynomial lattices]{Computing shortest vector in polynomial lattices}

We define polynomial lattices, some associated quantities and a few results that will be useful for us. In particular, we will discuss Alekhnovich's shortest vector computation algorithm in polynomial lattices.

\begin{definition}\label{defn:lattices}
  Let $B = \{ \beta_1, \beta_2, \dotsc, \beta_{\ell} \}$ be a set of $m$-dimensional vectors of polynomials, that is, $\beta_i \in \F[x]^m$ for all $i$. The \emph{lattice} $\mathcal{L}_B$ generated by this set over the ring $\F[x]$ is the set of all linear combinations 
  \[ \mathcal{L}_B := \left\{ f_i \beta_1 + f_2 \beta_2 + \dotsb + f_{\ell} \beta_{\ell} \colon f_i \in \F[x]\right\} .\]
  
  A \emph{basis} of a lattice is a set $B$ of  vectors of polynomials such that $B$ generates the lattice in the sense of the above definition and the vectors in $B$ are linearly independent over the field $\F(x)$ of rational functions over $\F$. 
\end{definition}
Before proceeding further, we make the following remark regarding the definition of lattices above. 
\begin{remark}
Lattices, as defined above are just finitely generated free modules over the univariate polynomial ring $\F[x]$. However, we stick to the term \emph{lattices} as opposed to free $\F[x]$-module. This is partly to be consistent with the convention in the work of Alekhnovich \cite{Alekhnovich2005}, and partly because of the similarities between some of the properties of finitely generated free modules over the univariate polynomial ring $\F[x]$ those of lattices over the ring of integers. 

We also note that for the results of this paper, it is important that we are working with finitely generated free modules over the univariate polynomial ring, and not the polynomial ring on a larger number of variables.  
\end{remark}

Notice that a basis $B$ can be thought of as an $m \times \ell$ matrix (the $\ell$ basis vectors written side by side as columns). When we refer to $\det B$ for a basis $B$ we are referring to the determinant of this matrix. As is the case with lattices over integers, this quantity is independent of which basis we choose for the lattice $\mathcal{L}$, i.e, if $B$ and $B'$ are two bases for the lattice $\mathcal{L}$, then $\det B= \det B'$, which we refer to as the determinant of the lattice $\mathcal{L}$, denoted by $\det \mathcal{L}$.

The following is the measure by which we define and seek a shortest vector in the lattice.

\begin{definition}\label{defn:norm}
  Let $\beta = (\beta(1),\beta(2), \dots, \beta(m))\in \F[X]^m$ be a vector of polynomials. We define \[ \deg \beta := \max_{j \in [m]} \deg \beta(j)\,. \]
  The \emph{leading coordinate} of $\beta$ is the largest coordinate where this maximum occurs, that is, \[ LC(\beta) := \max \{ j \colon \deg \beta(j) = \deg \beta \}\,.\qedhere \] 
\end{definition}
The notion of degree above gives us a way of comparing two vectors in $\F[x]^m$, which we informally refer to as the \emph{degree-norm}. We now define the notion of a reduced basis.

\begin{definition}\label{defn:basis and reduced basis}
  A \emph{reduced basis} $B$ of a lattice over $\mathcal{L}_B$ over the ring $\F[x]$ is one where the leading coordinates are all distinct. That is, for any pair of distinct non-zero elements $\beta_1, \beta_2 \in B$, we have $LC(\beta_1) \neq LC(\beta_2)$.
\end{definition}

Predictably, certain computations become easier when working with a reduced basis. However, it is unclear if such a basis exists for every lattice. The next proposition, due to Alekhnovich asserts that this is indeed true, and in fact he gives an algorithm to transform a general basis to a reduced one.

\begin{proposition}[{\cite[Proposition 2.1]{Alekhnovich2005}}]
Let $\mathcal{L}$ be a lattice  over the univariate polynomial ring $\F[x]$. Then, there exists a basis $B$ for $\mathcal L$ that is reduced as per \autoref{defn:basis and reduced basis}.
\end{proposition}
With a reduced basis at hand, the following observation of Alekhnovich gives a very simple algorithm to find a shortest vector. 
 
\begin{proposition}[{\cite[Proposition 2.3]{Alekhnovich2005}}]
Let $B$ be a reduced basis of a lattice $\mathcal L$ over the ring $\F[x]$. Let $\beta$ be any nonzero vector of minimal degree in $B$. Then $\beta$ is a shortest vector in $\mathcal L$. 
\end{proposition}

We will also be using a specific form of Minkowski's theorem guaranteeing the existence of a short vector in a lattice. The proof is simple and we give it here.

\begin{theorem}[Polynomial version of Minkowski's theorem]\label{thm:minkowski}
Let $\mathcal{L}$ be an $m$-dimensional polynomial lattice. Then, there is a nonzero vector $v$ satisfying
\[ \deg v \leq \frac{1}{m} \deg \det \mathcal{L} \] where $\deg v$ is the max-degree norm and $\det \mathcal{L}$ is the determinant of the lattice basis.
\end{theorem}

\begin{proof}
From above, using elementary matrix operations, we can assume that the basis matrix is lower triangular and has all the maximum degree elements on the diagonal. Then, the degree of the determinant is simply the sum of the degrees on the diagonal.

Since it is a reduced basis, by the above proposition, the shortest vector in the lattice is no shorter than the shortest basis vector. The shortest basis vector has degree at most the average of the degrees of the basis vectors, which we have ensured to occur on the diagonal. Since the sum of the degrees equals $\deg \det \mathcal{L}$, the average of the degrees is $(\deg \det \mathcal{L})/m$. 
\end{proof}

Finally, we state Alekhnovich's theorem pertaining to the fast algorithm for finding the shortest vector, that we later use off the shelf. The algorithm proceeds via a basis reduction.

\begin{theorem}[{\cite[Theorem 2.1]{Alekhnovich2005}}]\label{thm:alekhnovich shortest vector}
  Let $B$ be a set of $\ell$ vectors of dimension $m$. Let $n$ be the maximal degree of the polynomials which are entries of the vectors in $B$. Then, there is an algorithm that finds the shortest vector in $\mathcal{L}_B$, the lattice generated by $B$, in time $\tilde{O}(n(\ell + m)^4)$.
\end{theorem}

Note that the above theorem yields a nearly-linear-time algorithm in the maximal degree of the polynomials in the basis, albeit a polynomial time in the dimension of the lattice. Alekhnovich's key observation is that this suffices to obtain nearly-linear-time algorithms for several applications, one of which is list-decoding polynomial codes.

\section{Fast construction of the differential equation}\label{sec:mult-differential-eqn-const}
We start with the following definition that will be crucially used in our proof. In this form, this operator $\tau$ defined below and its variants show up naturally in the prior results on list decoding of multiplicity codes, e.g., \cite{GuruswamiW2013,BhandariHKS2024-mgrid}. One additional piece of notation  that we use throughout this section that for a polynomial univariate $A(x)$, and non-negative integer $k$, we use $A^{(k)}$ to denote $\frac{d^kA}{dx^k}$. 

\begin{definition}[Iterated derivative operator]\label{defn:differential-operator-for-derivative}
Let $s \in \N$ be a parameter. Then, for every $m$ such that $0 \leq m < s$, the function $\tau$ is an $\F$-linear map from the set of polynomials in $\F[x, y_0, y_1, \ldots, y_m]$ with $\vecy$-degree $1$ to the set of polynomials in $\F[x, y_0, y_1, \ldots, y_{m+1}]$ with $\vecy$-degree $1$ and is defined as follows. 
\[
\tau\left(\tilde{Q}(x) + \sum_{i = 0}^m Q_i(x)\cdot y_i\right) := \tilde{Q}^{(1)}(x) + \sum_{i = 0}^m \left( Q_i^{(1)}(x) \cdot y_i+ Q_i(x)\cdot y_{i+1}\right) \, .
\]

For an integer $i \leq (s-m) $, we use $\tau^{(i)}$ to denote the linear map from $\F[x, y_0, \ldots, y_m]$ to $\F[x, y_0, \ldots, y_{m + i}]$ obtained by applying the operator $\tau$ iteratively $i$ times.
\end{definition}
The operator $\tau$ is defined in this strange fashion for the following reason. For any polynomial $f \in \F[x]$, consider the univariate polynomial $P(x) := Q(x, f, f^{(1)}(x), \ldots, f^{(m)}(x))$. The derivative of $P$ with respect to $x$, namely $\frac{dP}{dx}$, is precisely equal to the polynomial obtained by evaluating $\tau(Q)$ on $(x, f, \ldots, f^{(m+1)})$.

We now state the main theorem of this section. 
\begin{theorem}\label{thm:multivariate-interpolation}
Let $R = (\alpha_i, \beta_{i, 0}, \ldots, \beta_{i, s-1})_{i = 1}^n$ be a received word for a multiplicity code decoder, and let $m \leq s$ be any parameter. 

Then, for $D \leq ((n(s-m))/m)$, there exists a non-zero polynomial $Q(x, y_0, y_1, \ldots, y_m)$ of the form $Q = \tilde{Q}(x) + \sum_i Q_i(x) \cdot y_i$, with degree of $\tilde{Q}$ and each $Q_i$ being at most $D$ such that for all $i \in \{0, 1, \ldots, s-m-1\}$,
\[
\tau^{(i)}(Q)(\alpha_i, \beta_{i, 0}, \ldots, \beta_{i, s-1}) = 0 \, .
\]

Moreover, there is a deterministic algorithm that when given $R$ as input, runs in time $\tilde{O}(n\poly(sm))$ and outputs such a polynomial $Q$ as a list of coefficients. 
\end{theorem}

\subsection{Proof of \autoref{thm:multivariate-interpolation}}
The proof of \autoref{thm:multivariate-interpolation} follows the high-level strategy in the work of Alekhnovich \cite{Alekhnovich2005} and proceeds via setting up the problem as a question of finding the shortest vector in an appropriate lattice over the ring $\F[x]$ and then applying the nearly-linear-time algorithm of Alekhnovich to find such a vector. The upper bound on the $x$-degree of $Q$ can once again be viewed as the consequence of properties of this lattice. We start by building the machinery necessary for the proof. 

\begin{definition}\label{def:lattice-definition}
Let $R = (\alpha_i, \beta_{i, 0}, \ldots, \beta_{i, s-1})_{i = 1}^n$ be a received word for a multiplicity code decoder, and let $m \leq s$ be any parameter (a positive integer). 

For $i \in \{0, 1, \ldots, m\}$, let $A_i(z)$ be the unique univariate polynomial of degree less than $n(s-i)$ in $\F[z]$ such that for every $j \in \{1, 2, \ldots, n\}$ and $k \in \{0, ,1, \ldots, s - i - 1\}$, 
\[
A_i^{(k)}(\alpha_j) = \beta_{j, i + k} \, .
\]
Let $\mathcal{L}_R^{(m)}$ be the $(m+2)$-dimensional lattice over the ring $\F[x]$ defined as 
\[
\mathcal{L}_R^{(m)} := \left\{\tilde{h}(x) \cdot \prod_{i = 1}^n (x - \alpha_i)^{s-m} + \sum_{i = 0}^m h_i(x)\cdot (y_i - A_i(x)) \colon \tilde{h}, h_i \in \F[x] \right\}\,.\qedhere
\]
\end{definition} 
One crucial bit of fact that we use later in our proof is that the polynomials $A_0, A_1, \ldots, A_m$ can be computed quite fast, given the received word as an input. More precisely, we have the following theorem that constructs a univariate polynomial from its high multiplicity evaluation given at a set of points, also known as \emph{Hermite interpolation}. We follow the notation in \autoref{def:lattice-definition}.
\begin{theorem}[{\cite[Algorithm 10.22]{GathenG-MCA}}]\label{thm:fast-hermite-interpolation}
Let $m \in \N$ be a parameter with $m \leq s$. Then, there is a deterministic algorithm that when given the received word $R = (\alpha_i, \beta_{i, 0}, \ldots, \beta_{i, s-1})_{i = 1}^n$ as input, outputs the polynomials $A_0, A_1, \ldots, A_m$ defined in \autoref{def:lattice-definition} in time $\tilde{O}(nsm)$. 
\end{theorem}

The following lemma summarizes some of the crucial properties of the lattice defined in \autoref{def:lattice-definition} and is the main technical step towards proving  \autoref{thm:multivariate-interpolation}.
\begin{lemma}\label{lem:properties-of-lattice}
Let $R = (\alpha_i, \beta_{i, 0}, \ldots, \beta_{i, s-1})_{i = 1}^n$ be a received word for a multiplicity code decoder, and let $m \leq s$ be any parameter, and let $\mathcal{L}_R^{(m)}$ be the lattice as defined in \autoref{def:lattice-definition}. Then, the following are true. 
\begin{itemize}
\item Any non-zero polynomial $Q$ in $\mathcal{L}_R^{(m)}$ is of the form $Q(x, y_0, y_1, \ldots, y_m) = \tilde{Q}(x) + \sum_{\ell} Q_{\ell}(x) \cdot y_{\ell}$ and for 
all $i \in \{0, 1, \ldots, s-m-1\}$,
\[
\tau^{(i)}(Q)(\alpha_i, \beta_{i, 0}, \ldots, \beta_{i, s-1}) = 0 \, .
\]
\item There is a non-zero polynomial in $\mathcal{L}_R^{(m)}$ with $x$-degree at most $D \leq (n(s-m)/m)$. 
\item Such a polynomial can be found in time $\tilde{O}(n\poly(sm))$.
\end{itemize}

\end{lemma}
\begin{proof}
We prove the lemma one item at a time, starting with the first item which describes the structure of polynomials in the lattice. 
\paragraph{Structure of polynomials in the lattice and behaviour under $\tau$: }Let $Q$ be a nonzero polynomial in $\mathcal{L}_R^{(m)}$. Thus, there are polynomials $\tilde{h}$ and $h_0, h_1, \ldots, h_m$ in $\F[x]$ such that 
\[
Q = \tilde{h}(x) \cdot \prod_{\ell = 1}^n (x - \alpha_{\ell})^{s-m} + \sum_{\ell = 0}^m h_{\ell}(x)\cdot (y_{\ell} - A_{\ell}(x)) \, .
\]
By separating out the terms containing the $y$ variables, we immediately get that $Q$ can be written as 
\[
Q(x, y_0, y_1, \ldots, y_m) = \tilde{Q}(x) + \sum_{\ell} Q_{\ell}(x) \cdot y_{\ell} \, ,
\] 
where for each $\ell$, $Q_{\ell} = h_{\ell}$ and $\tilde{Q}$ equals $\tilde{h}(x) \cdot \prod_{\ell = 1}^n (x - \alpha_{\ell})^{s-m} - \sum_j h_jA_j$.

To prove the second part of the first item in the lemma, we now try to understand the behaviour of $Q$ under the action of the operator $\tau$. In this context, the following two simple claims turn out to be useful. We first use the claims to prove the lemma, and then include their proofs. 

\begin{claim}\label{clm:h-tilde-under-tau}
For every $i \in \{0, 1, \ldots, s-m-1\}$ and $j \in \{1, 2, \ldots, n\}$, we have that the polynomial $\tau^{(i)}(\tilde{h}(x) \cdot \prod_{\ell = 1}^n (x - \alpha_{\ell})^{s-m})$ when evaluated at $(\alpha_j, \beta_{j, 0}, \ldots, \beta_{j, s})$ is zero. 
\end{claim}

\begin{claim}\label{clm:yi-Ai-under-tau}
For every $i \in \{0, 1, \ldots, s-m-1\}$, $j \in \{1, 2, \ldots, n\}$ and $\ell \in \{0, 1, \ldots, m\}$, we have that the polynomial $\tau^{(i)}(h_{\ell}(x)(y_{\ell} - A_{\ell}(x)))$ when evaluated at $(\alpha_j, \beta_{j, 0}, \ldots, \beta_{j, s})$ is zero.
\end{claim}
From the definition of the operator $\tau$, we get that it is a linear map on the polynomial ring, and thus for every $i \in \N$, $\tau^{(i)}$ is also a linear map on the polynomial ring. Thus, for every $i \in \{0, 1, \ldots, s-m-1\}$, $\tau^{(i)}(Q)$ equals 
\[
\tau^{(i)}(Q) = \tau^{(i)}\left(\tilde{h}(x) \cdot \prod_{\ell = 1}^n (x - \alpha_{\ell})^{s-m}\right) + \sum_{\ell = 0}^m \tau^{(i)}\left(h_{\ell}(x)\cdot (y_{\ell} - A_{\ell}(x)) \right) \, .
\]
From \autoref{clm:h-tilde-under-tau} and \autoref{clm:yi-Ai-under-tau}, it follows that each of the summands above vanishes when evaluated at $(\alpha_j, \beta_{j, 0}, \ldots, \beta_{j, s})$ for every $j \in \{1, \ldots, n\}$. This completes the proof of the first item in the lemma. 

\paragraph{Existence of a low-degree polynomial in the lattice: }
To show that there is a low-degree polynomial in the lattice ${\cal L}_R^{(m)}$, we note that while we have been viewing polynomials in the lattice as polynomials of degree at most $1$ in the variables $y_0, y_1, \ldots, y_m$ with coefficients in $\F[x]$, we can equivalently view any such polynomial  $Q$ as an $(m + 2)$-dimensional column vector with entries in the ring $\F[x]$, where the coordinates are labelled $(0, 1, \ldots, m, m+1)$ and for $\ell < m+1$, the $\ell^{th}$ coordinate corresponds to the coefficient of $y_{\ell}$ in the polynomial $Q$, and the coordinate labelled $(m+2)$  stores the $y$-free part of $Q$. For instance, if $Q = \tilde{Q} + \sum_{\ell=0}^m h_{\ell}$, then this corresponds to the transpose of $(Q_0, Q_1, \ldots, Q_m, \tilde{Q})$. Moreover, we note that the vectors corresponding to the generators of the lattice ${\cal L}_R^{(m)}$, namely the polynomials $(y_0 - A_0), \ldots, (y_m - A_m)$ and $\prod_{\ell = 1}^n (x - \alpha_{\ell})^{s-m}$ form a lower triangular matrix with non-zero diagonal entries, and hence are linearly independent over $\F(x)$. In other words, the lattice ${\cal L}_R^{(m)}$ is full rank. Moreover, all but one diagonal entries is just $1$ and the bottom-most diagonal entry is precisely $\prod_{\ell = 1}^n (x - \alpha_{\ell})^{s-m}$, which is a polynomial of degree $n(s-m)$. Thus, the determinant of this matrix is a polynomial of degree at most (in fact, it is equal to) $n(s-m)$. From Minkowski's Theorem (\autoref{thm:minkowski}), we have that there is a vector in the lattice ${\cal L}_R^{(m)}$ with every entry of degree at most $n(s-m)/(m+2) \leq n(s-m)/m$. This completes the proof of the second item in the lemma\footnote{The curious reader familiar with previous list-decoding algorithms for polynomial codes \cite{Sudan1997,GuruswamiS1999,Alekhnovich2005,GuruswamiR2008,Kopparty2014,GuruswamiW2013,BhandariHKS2024-mgrid,BhandariHKS2024-affineFRS} will notice that the argument presented here is different from the standard argument which proceeds by constructing a homogenous system of linear equations and showing that a non-zero solution exists if the number of variables exceeds the number of constraints. The alternate argument presented here using Minkowski's Theorem is inspired by a similar argument due to Coppersmith \cite{Coppersmith1997} in the context of finding small integer solutions to polynomial equations.}. 

\paragraph{Constructing a low-degree polynomial efficiently: }
In order to construct a shortest vector in the lattice ${\cal R}$, we invoke the algorithm of Alekhnovich in \autoref{thm:alekhnovich shortest vector} that takes the set of generators for the lattice as input, and outputs a shortest vector in time $\tilde{O}(n\cdot \poly(sm))$. Here, we use the fact that the lattice is given by $m+2$ vectors in dimensions $m+2$ over the ring $\F[x]$ and each entry of the vectors is a polynomial of degree at  most $n(s-m)$ and thus two such polynomials can be multiplied in time at most $\tilde{O}(n(s-m))$.  
\end{proof}
Now, to complete the proofs of the lemma, we sketch the proofs of \autoref{clm:h-tilde-under-tau} and \autoref{clm:yi-Ai-under-tau}.
\begin{proof}[Proof of \autoref{clm:h-tilde-under-tau}]
On  a polynomial that just depends on $x$ variables, the operator $\tau$ operates by just differentiating it with respect to $x$ once. In particular, this implies that for any $i \in \{0, 1, \ldots, s-m-1\}$, and any polynomial $\tilde{h} \in \F[x]$, $\tau^{(i)}(\tilde{h} \cdot \prod_{\ell = 1}^m (x-\alpha_{\ell})^{s-m})$ must be divisible by $\prod_{\ell = 1}^m (x-\alpha_{\ell})$,and thus vanishes when the variable $x$ is set to a value in $\{\alpha_1, \ldots, \alpha_n\}$.  
\end{proof}

\begin{proof}[Proof of \autoref{clm:yi-Ai-under-tau}]
For the proof of this claim, the key observation is that for every $i < s-m$, the polynomial $\tau^{(i)}(h_{\ell}\cdot (y_{\ell} - A_{\ell}))$ is of the form $\sum_{k = 0}^i g_k(x)\cdot (y_{\ell + k} - A_{\ell}^{(k)})$ for some polynomials $g_0, g_1, \ldots, g_i \in  \F[x]$. This can be seen by the definition of the operator $\tau$ and an induction on $i$. For $i = 0$, this is vacuously true. Inductively, we assume that $i < s-m$ and $\tau^{(i-1)}(h_{\ell}\cdot (y_{\ell} - A_{\ell}))$ is of the form $\sum_{k = 0}^{i-1} g_k(x)\cdot (y_{\ell + k} - A_{\ell}^{(k)})$. Now, applying $\tau$ to this expression one more time (and we know that this can be done since $i < s-m$), we get 
\[\tau^{(i)}(h_{\ell}\cdot (y_{\ell} - A_{\ell})) = \tau \left(\sum_{k = 0}^{i-1} g_k(x)\cdot (y_{\ell + k} - A_{\ell}^{(k)})\right) \, .
\]
By definition of $\tau$, this gives 
\[
\tau^{(i)}(h_{\ell}\cdot (y_{\ell} - A_{\ell})) =  \left(\sum_{k = 0}^{i-1} \left( g_k^{(1)}\cdot (y_{\ell + k} - A_{\ell}^{(k)}) + g_k\cdot (y_{\ell + k + 1} - A_{\ell}^{(k+1)})\right)\right) \, .
\]
Thus, by reindexing the summations, we have the desired structure. 

Once we have this structure, we note that evaluating any polynomial of the form $\sum_{k = 0}^i g_k(x)\cdot (y_{\ell + k} - A_{\ell}^{(k)})$ at $(\alpha_j, \beta_{j, 0}, \ldots, \beta_{j, s})$ gives us $\sum_{k = 0}^i g_j(\alpha_j)\cdot (\beta_{j, \ell + k} - A_{\ell}^{(k)}(\alpha_j))$.
But from the interpolation condition on $A_{\ell}$ in \autoref{def:lattice-definition}, we have that for every $j\in \{1, 2, \ldots, n\}$, $(\beta_{j, \ell + k} - A_{\ell}^{(k)}(\alpha_j))$ equals zero. This completes the proof of the claim.  
\end{proof}
We are now ready to complete the proof of \autoref{thm:multivariate-interpolation}. 

\begin{proof}[Proof of \autoref{thm:multivariate-interpolation}]
Given the received word $R$, we construct the basis for the lattice ${\cal L}_R^{(m)}$ in \autoref{def:lattice-definition}. This basis can be constructed in time at most $\tilde{O}(n\poly(sm))$ since all the $A_i$'s can be constructed in this time from \autoref{thm:fast-hermite-interpolation} and the polynomial $\prod_{i=1}^n(x-\alpha_i)^{s-m}$ can also be constructed in this time using the Fast Fourier Transform. We now invoke \autoref{lem:properties-of-lattice} and get that there is a non-zero polynomial $Q(x, y_0, y_1, \ldots, y_m)$ in ${\cal L}_R^{(m)}$ of the form $Q = \tilde{Q} + \sum_{i = 0}^m Q_i(x)y_i$ that can be constructed in time $\tilde{O}(n\poly(sm))$ such that the $x$-degree of $Q$ is at most $D \leq (n(s-m)/m)$ and it satisfies precisely the linear constraints with respect to $R$ that the theorem seeks. 
\end{proof}

\subsection{Close enough codewords satisfy the equation}
We now prove the following simple lemma that shows that every message polynomial $f$ whose encoding is close enough to the received word $R$ \emph{satisfies} the differential equation $Q$. This step is identical to that in \cite{GuruswamiW2013}.
\begin{lemma}\label{lem:close-enough-codewords-satisfy-the-differential-equation}
Let $R = (\alpha_i, \beta_{i, 0}, \ldots, \beta_{i, s-1})_{i = 1}^n$ be a received word for a multiplicity code decoder, $m \leq s$ be a parameter, and let $Q(x, y_0, y_1, \ldots, y_m) \in \mathcal{L}_R^{(m)}$ be a non-zero polynomial with $x$-degree at most $D \leq (n(s-m)/m)$.  

Then, for every polynomial $f \in \F[x]$ of degree at most $d$ such that $\enc_{s, \mathbb{\alpha}}(f)$ agrees with $R$ on more than $(D + d)/(s-m)$ values of $i$, $Q(x, f, f^{(1)}(x), \ldots, f^{(m)}(x))$ is identically zero. 

\end{lemma}
\begin{proof}
Consider the univariate polynomial $P(x) = Q(x, f, f^{(1)}(x), \ldots, f^{(m)}(x))$. The degree of $P$ is at most $D + d \leq d + n(s-m)/m$. Recall that the $\tau$ operator was defined such that the derivative of $P$ with respect to $x$, $\frac{dP}{dx}$ is precisely equal to the polynomial obtained by evaluating $\tau(Q)$ on $(x, f, \ldots, f^{(m+1)})$.  More generally, for every $i \in \{0, 1, \ldots, s-m-1 \}$, we have that $\frac{d^{i}P}{dx^i}$ is precisely equal to the univariate polynomial obtained by evaluating  $\tau^{(i)}(Q)$ on the input $((x, f, \ldots, f^{(m+i)})$. Thus, for any $j \in \{1, 2, \ldots, n\}$, if the $j^{th}$ coordinate of the encoding of $f$ agrees with the received word $R$, i.e.,$(\alpha_j, f(\alpha_j), \ldots, f^{(s-1)}(\alpha_j))$ equals $(\alpha_j, \beta_{j, 0}, \ldots, \beta_{j, s-1})$, then, from the above discussion, we have that 
\[
\frac{d^{i}P}{dx^i}(\alpha_j) = \tau^{(i)}(Q)(\alpha_j, f(\alpha_j), \ldots, f^{(s-1)}(\alpha_j)) = \tau^{(i)}(Q)(\alpha_j, \beta_{j, 0}, \ldots, \beta_{j, s-1}) \, ,
\]
which by the constraints imposed on $Q$ in the definition of ${\cal L}_R^{(m)}$ is zero. 

Thus, for every point of agreement between $R$ and the encoding of $f$, $P$ vanishes with multiplicity at least $(s-m)$. Therefore, if the number of such agreements exceeds $(D + d)/(s-m) \leq d/(s-m) + n/m$, we have that $P$ must be identically zero.  
\end{proof}

\section{Solving the differential equation in nearly-linear time}\label{sec:solving-differential-equations}
In this section, we design fast algorithms for \emph{solving} differential equations of the form \[Q\left(x, f, \ldots, f^{(m)}\right) \equiv 0\] obtained in \autoref{lem:close-enough-codewords-satisfy-the-differential-equation}. Since the degree of $Q$ in $\vecy$ variables is at most $1$, we have that the solution space of all polynomials of degree at most $d$ satisfying $Q(x, f, \ldots, f^{(m)}) = 0$ is an affine space. It was shown by Guruswami \& Wang \cite{GuruswamiW2013} that this space has dimension at most $m$ and a basis for this space can be obtained in deterministic polynomial time. In the main theorem of this section, stated below, we show that such a basis can, in fact be obtained significantly faster.  
\begin{theorem}\label{thm:fast-differential-equation-solver-original}
Let $\F$ be a finite field of characteristic greater than $d$ or zero, and let \[Q(x, y_0, \ldots, y_m) = \tilde{Q}(x) + \sum_{i = 0}^m Q_i(x)\cdot y_i\] be a non-zero polynomial with $x$-degree at most $D$. Then, the affine space of polynomials $f(x) \in \F[x]$ of degree at most $d$ that satisfy 
\[
Q\left(x, f, f^{(1)} \ldots, f^{(m)}\right) \equiv 0
\]
has dimension at most $m$.

  Moreover, there is a deterministic algorithm that when given $Q$ as an input via its coefficient vector, and the parameter $d$, performs at most $\tilde{O}((D+d)\poly(m))$ field operations and outputs a basis  for this affine space. 
\end{theorem}

Since $Q$ is a non-zero polynomial, it follows that at least one of the polynomials $\tilde{Q}, Q_0, \ldots, Q_m$ is non-zero. If $\tilde{Q}$ is the only non-zero polynomial amongst these, then the space of solutions for $Q(x, f, f^{(1)}(x), \ldots, f^{(m)}(x)) \equiv 0$ is clearly empty, and this condition can clearly be checked in linear time since $Q$ is given via its coefficient vector. Therefore, for the rest of this discussion, we assume that at least one of the $Q_i$'s is non-zero. Moreover, let $m_0$ be the largest $i \in \{0, 1, \ldots, m\}$ that is non-zero. Thus, $Q$ is only a function of $x$ and $y_0, y_1, \ldots, y_{m_0}$ in this case. For simplicity of notation, we just assume that $m_0 = m$, i.e., $Q_{m}(x)$ is a non-zero polynomial.

For further ease of notation, we also assume in the rest of this section that the constant term of the polynomial $Q_m(x)$, i.e.,$Q_m(0)$ is non-zero. Since $Q_m$ is assumed to be non-zero, there is a constant $\beta$ in $\F$ (or an extension of $\F$ if the size of $\F$ is less than the degree $D$ of $Q_m$) such that $Q_m(\beta)$ is non-zero. Up to a translation $x \to x + \beta$ of the $x$ variable, $\beta$ can also be assumed to be the origin, and thus, for the rest of this section, we assume without loss of generality that $Q_m(0)$ is non-zero. 

Based on the above discussion, in order to prove \autoref{thm:fast-differential-equation-solver-original} it suffices to prove the following theorem. 
\begin{theorem}\label{thm:fast-differential-equation-solver}
Let $\F$ be a finite field of characteristic greater than $d$ or zero, and let \[Q(x, y_0, \ldots, y_m) = \tilde{Q}(x) + \sum_{i = 0}^m Q_i(x)\cdot y_i\] such that its $x$-degree is at most $D$ and $Q_m(0)$ is non-zero. Then, the affine space of polynomials $f(x) \in \F[x]$ of degree at most $d$ that satisfy 
\[
Q\left(x, f, f^{(1)} \ldots, f^{(m)}\right) \equiv 0
\]
has dimension at most $m$.

  Moreover, there is a deterministic algorithm that when given $Q$ as an input via its coefficient vector, and the parameter $d$, performs at most $\tilde{O}((D+d)\poly(m))$ field operations and outputs a description (at most $m+1$ vectors whose affine span is this space) for this affine space. 
\end{theorem}

\subsection{Set up for proof of \autoref{thm:fast-differential-equation-solver}}  
As we alluded to in the previous section, the bound of $m$ on the dimension of the affine space of solutions of degree at most $d$ of $Q\left(x, f, f^{(1)} \ldots, f^{(m)}\right) \equiv 0$ was already shown by Guruswami and Wang \cite{GuruswamiW2013}. They showed the following. 
 \begin{theorem}[\cite{GuruswamiW2013}]\label{thm:dim-of-solution-space-guruswami-wang}
Let $k$ be any natural number, $\F$ be a finite field of characteristic at least $k+m$ or zero, and let \[Q(x, y_0, \ldots, y_m) = \tilde{Q}(x) + \sum_{i = 0}^m Q_i(x)\cdot y_i\] be such that its $x$-degree is at most $D$ and $Q_m(0)$ is non-zero. Then, the set of polynomials $f$ of degree at most $(k+m-1)$ that satisfy \[Q\left(x, f, f^{(1)} \ldots, f^{(m)}\right) \equiv 0  \mod x^k \, .\] form an affine space of dimension  $m$. 

Moreover, there is a $\poly(k, D, m)$ time algorithm that when given the coefficient vector of $Q$ and the parameter $d$ as inputs, outputs a basis for this affine space. 


\end{theorem}  

The proof in Guruswami \& Wang \cite{GuruswamiW2013} can be viewed as an application of the widely used power series based techniques for solving differential equations, and essentially shows that given the coefficients of $x^0, x, x^2, \ldots, x^{m-1}$ in any solution $f$ of this differential equation, there is a \emph{unique} way of recovering the remaining coefficients. In order to improve the computational efficiency of this algorithm and prove \autoref{thm:fast-differential-equation-solver}, we argue that this iterative procedure can be implemented faster than just recovering one coefficient at a time. 

Our high level idea is essentially a reduction from looking for degree $d$ solutions for this equation to looking for equations of degree close to $d/2$ of two closely related equations with appropriate structure, which are then recursively solved. In spirit, this is similar to the well known Newton Iteration/Hensel Lifting based algorithm that computes the inverse of a given univariate polynomial\footnote{Perhaps interestingly, the input polynomial here is also assumed to have a non-zero constant term. This is essentially for the same reason that we have assumed (without loss of generality here) that $Q_m$ has a non-zero constant term.} modulo $x^d$ in nearly-linear time (e.g., see \cite[Theorem 9.4]{GathenG-MCA}), except for one key difference: in the aforementioned fast algorithm for modular inverse computation in \cite{GathenG-MCA}, the two instances of roughly half the size generated in the recursive call are precisely the same. So, effectively, in every recursive call, there is some nearly-linear-time computation before and after the call, and precisely one recursive call on a problem instance of roughly half the size. Unfortunately, for our algorithm, the overall structure is not so clean. Things are also a little more complicated by the fact that unlike for inverse computation,the solution to the equations we are hoping to solve are not really unique. However, as it turns out, the smaller recursive calls are to problem instances that  continue to have some additional structure that lets us make this outline work. 

We start by describing the kind of differential equations that make an appearance in our recursive calls. But first we set up some necessary notation. For brevity of notation, we denote a polynomial $A\left(x, f, f^{(1)} \ldots, f^{(m)}\right)$ by $A(x, \partial f)$. Throughout, the field $\F$ and the polynomial $Q, \tilde{Q}, Q_0, \ldots, Q_m$ are as given in the hypothesis of \autoref{thm:fast-differential-equation-solver}, and in particular the constant term of $Q_m$ is non-zero. The following is a crucial definition. 
\begin{definition}\label{def:conjugate}
For a polynomial $Q(x, \vecy) = \tilde{Q}(x) + \sum_{i = 0}^m Q_i(x)\cdot y_i$, and an $n \in \N$, we use $Q^{\dagger}_n$ to denote the polynomial 
\[
Q^{\dagger}_n(x,\vecy) := \sum\limits_{i=0}^{m} x^{i}\left(\sum\limits_{j=i}^{m}\frac{n!}{(n+i-j)!}\binom{j}{i}\cdot x^{m-j}\cdot Q_{j}(x)\right)\cdot y_i\,.\qedhere
\]
\end{definition}
The polynomial $Q^{\dagger}$ is defined in this strange fashion as it satisfies the property that for any $m$, we have $Q_m^{\dagger}(x,\partial f) = \sum_{i=0}^m Q_i(x)\cdot (x^m \cdot f)^{(i)}$, a property we will prove and use later. Before further elaborating on the role \autoref{def:conjugate} has to play in our proof of \autoref{thm:fast-differential-equation-solver}, we  state the following lemma that summarises the structure of the solution space of the differential equation $Q^{\dagger}_n(x, \partial f) \equiv 0$.
\begin{lemma}\label{lem:solution-space-dagger}
Let $n, k \in \N$ and $\Char(\F)$ is either zero or greater than or equal to $\max(n+k, m)$. Let $Q(x, \vecy) = \tilde{Q}(x) + \sum_{i = 0}^m Q_i(x)\cdot y_i \in \F[x, \vecy]$ be such that $Q_m(0)$ is non-zero. Then, for every $B(x) \in \F[x]$, the following are true. 
\begin{itemize}
\item The differential equation $\left(B(x) + Q^{\dagger}_n(x, \partial f) \equiv 0 \mod x^k\right)$ has a unique solution modulo $x^k$. 
\item For every $\ell \in \N$ with $\ell < k$, this unique $f(x)$ satisfying $\left( B(x) + Q^{\dagger}_n(x, \partial f) \equiv 0 \mod x^k\right)$  can be written as $f(x) = h(x) + x^{\ell}g(x)$ where, 
\begin{itemize}
\item $h$ is the unique polynomial (modulo $x^{\ell}$) that satisfies $\left(B(x) + Q^{\dagger}_n(x, \partial h) \equiv 0 \mod x^{\ell}\right)$ 
\item $g$ is the unique polynomial (modulo $x^{k-\ell}$) that satisfies $\left(\tilde{B}(x) + Q^{\dagger}_{n+\ell}(x, \partial g) \equiv 0 \mod x^{k-\ell}\right)$, where $\tilde{B}(x) := x^{-\ell}\cdot (B(x) + Q^{\dagger}_n(x, \partial h))$. \footnote{We recall that since $\left(B(x) + Q^{\dagger}_n(x, \partial h) \equiv 0 \mod x^{\ell}\right)$, $\tilde{B}$ as defined here is indeed a polynomial.} 
\end{itemize}
\end{itemize}
\end{lemma}
The relevance of the \autoref{def:conjugate} towards the proof of \autoref{thm:fast-differential-equation-solver} stems from the following lemma that drives our algorithm. 
\begin{lemma}\label{lem:solution-space-original}
Let $k$ be a natural number and $\Char(\F)$ is either zero or greater than or equal to $m+k$. Let $Q(x, \vecy) = \tilde{Q}(x) + \sum_{i = 0}^m Q_i(x)\cdot y_i \in \F[x, \vecy]$ be such that $Q_m(0)$ is non-zero. 

Then, for every $h \in \F[x]$ of degree at most $(m-1)$, there is a unique $g \in \F[x]$ of degree less than $k$ such that the polynomial  $f := h + x^m \cdot g$ is a solution of $Q(x, \partial f) \equiv 0 \mod x^k$.

Moreover, the polynomial $g$ is the unique solution of $B(x) + Q^{\dagger}_m(x, \partial g) \equiv 0 \mod x^k$, where $B(x) := Q(x, \partial h)$.
\end{lemma}
We recall from \autoref{thm:dim-of-solution-space-guruswami-wang} that solutions to $Q(x, \partial f) \equiv 0 \mod x^k$ form an affine space of dimension  $m$. Combining this bound on the dimension with \autoref{lem:solution-space-original}, we get the following. 
\begin{lemma}\label{lem:basis-of-solution-space-original}
Let $k$ be a natural number and $\Char(\F)$ be either zero or greater than or equal to $m+k$. Let $Q(x, \vecy) = \tilde{Q}(x) + \sum_{i = 0}^m Q_i(x)\cdot y_i \in \F[x, \vecy]$ be such that $Q_m(0)$ is non-zero. For $i \in \{0, 1, \ldots, m\}$ \footnote{We go up till $m$ as it works like plugging in $0$ instead of an $x^i$ and is cleaner notationally.}, let $g_i(x)$ be the unique polynomial of degree less than $k$ guaranteed by \autoref{lem:solution-space-original} such that $f_i := x^i + x^m\cdot g_i$ is a solution of  $Q(x, \partial f_i) \equiv 0 \mod x^k$.

Then, $f_0, f_1, \ldots, f_{m}$ form a basis of the affine space of solutions of degree at most $(m+k-1)$ of $Q(x, \partial f) \equiv 0 \mod x^k$. 
\end{lemma}
\begin{proof}[Proof sketch]
The lemma immediately follows from the fact that $f_0-f_m, f_1-f_m, \ldots, f_{m-1}-f_m$ are linearly independent over $\F$ and the dimension of the solution space of $Q(x, \partial f) \equiv 0 \mod x^k$ is $m$, as is given by \autoref{thm:dim-of-solution-space-guruswami-wang}.
\end{proof}
The basis of the solution space described by \autoref{lem:basis-of-solution-space-original} is the basis that we construct in our algorithm. We defer the proofs of \autoref{lem:solution-space-dagger} and \autoref{lem:solution-space-original} to the end of this section, and proceed with the discussion and analysis of our algorithms. We start with an algorithm that uses \autoref{lem:solution-space-dagger} to solve the equations that appear therein. 

\subsection{Algorithm for solving \texorpdfstring{$B(x) + Q^{\dagger}_n(x, \partial f) \equiv 0$}{B(x)+Qdagger(x, df) =0}}

\newcommand{\solveQdag}{\texttt{SolveQ$^{\dagger}$}\xspace}
\begin{minipage}{\algwidth}
\begin{algorithm}[H]
\caption{\solveQdag (Solving equations of the form $B(x) + Q^{\dagger}_n(x, \partial f) \equiv 0$)}\label{algo:solve-dagger}
\KwIn{
  $(Q(x, \vecy),B(x),n,k)$, where $Q(x, \vecy) \in \F[x, \vecy], B(x) \in \F[x]$ are polynomials with $Q(x, \vecy) = \tilde{Q}(x) + \sum_{i = 0}^m Q_i(x)\cdot y_i$ with $Q_m(0) \neq 0$, $x$-degree at most $D$, $n, k$ are natural numbers and $\Char{\F}$ is either zero or at least $n + k$. }
\KwOut{The unique polynomial $f(x) \in \F[x]$ of degree less than $k$ that satisfies $B(x) + Q^{\dagger}_n(x, \partial f) \equiv 0 \mod x^k$, where $Q^{\dagger}_n$ is defined as in \autoref{def:conjugate}.}
\nonl\hrulefill\\
\eIf{$k = 1$}{
	\KwRet{$-B(0)/{Q_m(0)}$ ;}
}
{ Set $\ell \gets \lfloor k/2 \rfloor$ ;\\
	Recursively run \solveQdag (\autoref{algo:solve-dagger}) on input $(Q,B,n,\ell)$ to obtain the polynomial $h(x)$ ;\\
  Recursively run \solveQdag (\autoref{algo:solve-dagger}) on input $\left(Q,x^{\ell}\cdot (B(x) + Q^{\dagger}_n(x,\partial h)),n+\ell,k-\ell\right)$ to obtain the polynomial $g(x)$ ;\\
	\KwRet{$h(x) + x^{\ell}\cdot g(x)$ .}
}
\end{algorithm}
\end{minipage}

 We now bound the running time of the algorithm, and prove its correctness. 
 \begin{lemma}[Time complexity of \solveQdag]\label{lem:algo-dagger-runningtime}
  \solveQdag (\autoref{algo:solve-dagger}) requires at most $\tilde{O}(k\poly(m))$ many field operations in the worst case.
 \end{lemma}
 \begin{proof}
 The algorithm is essentially a divide and conquer, where in order to obtain a solution of the original differential equation modulo $x^k$, we solve two instances of a related equation modulo $x^{k/2}$ if $k$ is even, and modulo $x^{\lfloor k/2 \rfloor}$, $x^{\lceil k/2 \rceil}$ if $k$ is odd. Moreover, the solutions $g, h$ of these smaller instances can be combined into a solution $h + x^{\lfloor k/2 \rfloor}\cdot g$ using at most $O(k)$ field operations. One additional bit of processing that is needed before the second recursive call is in constructing the polynomial $x^{-\ell}\cdot (B + Q^{\dagger}_n(x, \partial h))$. We note that we only need to compute this polynomial modulo $x^{\lceil k/2 \rceil}$, thus we can assume without loss of generality that $B$ as well as each of the $Q_i$s have degree at most $x^{\lceil k/2 \rceil}$. Now, given an $h$ of degree at most $\ell$, we can compute the polynomials $h, h^{(1)}, \ldots, h^{(m)}$ with at most $O(km)$ field operations using \autoref{lem:computing-derivatives-fast}. For $i, j \leq m$, computing the binomial coefficients $\binom{j}{i}$ takes at most $O(m)$ field operations by just naively expressing this as $\frac{j!}{i!(j-i)!}$. Similarly, computing $\frac{n!}{(n+i-j)!} = n(n-1)\cdots (n-i + j + 1)$ takes at most $O(m)$ field operations for every $n \in \N$. Thus, every term  $\left(\sum_{j = i}^m \frac{n!}{(n+i-j)!}\binom{j}{i}x^{m-j}Q_j\right)$ can be computed using at most $O(k\poly(m))$ field operations, since the degree of each $Q_i$ can be assumed to be at most $x^{\lceil k/2 \rceil}$. Finally, to compute 
 \[
B + \sum_{i = 0}^m x^i\left(\sum_{j = i}^m \frac{n!}{(n+i-j)!}\binom{j}{i}x^{m-j}Q_j\right)h^{(i)}  
 \,,\]
we need to additionally perform $m$ polynomial multiplications, $m$ summations of univariates of degree at most $x^{\lceil k/2 \rceil}$ and some shifts (corresponding to multiplications by monomials), and hence the overall complexity of computing  $x^{-\ell}(B + Q^{\dagger}_n(x, \partial h))$ is at most $O(k\poly(m))$ field operations.

Thus, the total number of field operations in any execution of the algorithm can be bounded via the recursion
\[
T(k) \leq 2T(\lceil k/2 \rceil) + O(k\poly(m)) \, ,
\]
  which immediately yields $T(k) \leq \tilde{O}(k\poly(m))$. 
 \end{proof}
 \begin{lemma}[Correctness of \solveQdag]\label{lem:algo-dagger-correctness}
For every input $(Q, B, n, k)$,  if the underlying field $\F$ of characteristic either zero or greater than or equal to $n+k$, \solveQdag (\autoref{algo:solve-dagger}) correctly outputs the unique solution of  $B + Q^{\dagger}_n(x, \partial f) \equiv 0 \mod x^k$.
 \end{lemma}
\begin{proof}
We note that as we make recursive calls in the algorithm, the sum of the parameters $n_i$ and $k_i$ for each call continues to be at most $n+k$. So, the condition on the characteristic of the field being either zero or greater than $n+k$ continues to hold at each recursive call. 

That the base case of the algorithm $(k = 1)$ outputs the correct solution is straightforward to see. The overall correctness now immediately follows from the second item of \autoref{lem:solution-space-dagger}. 
\end{proof}

    
    

\subsection{Algorithm for solving \texorpdfstring{$Q(x, \partial f) \equiv 0$}{Q(x,dx)=0}}

\newcommand{\solveQ}{\texttt{SolveQ}\xspace}
\begin{minipage}{\algwidth}
\begin{algorithm}[H]
\caption{\solveQ (Solving equations of the form $Q(x, \partial f) \equiv 0$)} \label{algo:solve-original}
\KwIn{$(Q(x, \vecy), d)$, where $Q(x, \vecy) \in \F[x, \vecy]$ is polynomials with $Q(x, \vecy) = \tilde{Q}(x) + \sum_{i = 0}^m Q_i(x)\cdot y_i$ with $Q_m(0) \neq 0$, $x$-degree $D$, $d$ is a natural number and $\Char{\F}$ is either zero or greater than ${d}$.}
\KwOut{A basis for the affine space of solutions of degree at most $d$ of $Q(x, \partial f) \equiv 0$. }

\For{$i \gets 0$ to $m$}{
  Run \solveQdag (\autoref{algo:solve-dagger}) on inputs $(Q,Q(x,\partial x^i),m,d-m+1)$ to obtain the polynomial $g_i$ ;\\
  Set $f_i \gets x^i + x^m \cdot g_i$ ;
}
\KwRet{$f_0,f_1,\dots,f_m$ .}
\end{algorithm}
\end{minipage}
 
 \begin{lemma}[Time complexity of \solveQ]\label{lem:algo-original-runningtime}
\solveQ (\autoref{algo:solve-original}) requires at most $\tilde{O}((D+d)\poly(m))$ many field operations in the worst case.
 \end{lemma}
 \begin{proof}
 The algorithm makes $m$ calls to \solveQdag (\autoref{algo:solve-dagger}), and hence its time complexity is at most $m$ times the complexity of the most expensive of these calls and the time taken for preprocessing before making the call. For the $i^{th}$ call, we need to construct the coefficient representation of the polynomial $Q(x, \partial x^i)$, which takes at most $O(D\poly(m))$ field operations. 
  
 Thus, using \autoref{lem:algo-dagger-runningtime}, we have that the overall complexity of the algorithm is at most $O(D\poly(m)) + \tilde{O}(d \poly(m)) \leq \tilde{O}((D+d)\poly(m))$ many field operations. 
 \end{proof}
 \begin{lemma}[Correctness of \solveQ]\label{lem:algo-original-correctness}
 For every input $Q, d$, if the underlying field has characteristic zero or greater than $d$, then \solveQ (\autoref{algo:solve-original}) correctly outputs a basis for the space of solutions of degree at most $d$ for $Q(x, \partial f) \equiv 0$.
 \end{lemma}  
 \begin{proof}
 Since we are looking for solutions of degree at most $d$ of $Q(x, \partial f) \equiv 0$, we have from \autoref{thm:dim-of-solution-space-guruswami-wang} that it suffices to solve $Q(x, \partial f) \equiv 0$ modulo $x^{d-m+1}$. 
 
From \autoref{lem:basis-of-solution-space-original}, we have that for $i = 0, 1, \ldots, m$, polynomials of the form $f_i := x^i + x^mg_i$ form a basis of the $m$-dimensional affine space of solutions of degree at most $d$ of $Q(x, \partial f) \equiv 0 \mod x^{d-m+1}$, where $g_i$ is the unique solution of degree at most $d$ of $Q(x, \partial x^i) + Q^{\dagger}_m(x, \partial g_i) \equiv 0 \mod x^{d-m+1}$. From \autoref{lem:algo-dagger-runningtime}, we have that each $g_i$ is computed correctly in the call to \solveQdag (\autoref{algo:solve-dagger}). The lemma now immediately follows from \autoref{lem:basis-of-solution-space-original}.
 \end{proof}

%
  

 \subsection{Proofs of technical lemmas \texorpdfstring{({\cref{lem:solution-space-dagger,lem:solution-space-original}})}{}}

\begin{proof}[Proof of \autoref{lem:solution-space-dagger}]
To show the uniqueness of solution of $B(x) + Q^{\dagger}_n(x, \partial f) \equiv 0 \mod x^k$, we view the equation as a linear system in the coefficients of $f$. We argue that the linear system we get is square, lower triangular and all the diagonal elements are non-zero, thereby implying that the solution is unique. 

We start by arguing that the system is lower triangular. Let $f = f_0 + f_1x + \cdots + f_dx^d$ be a polynomial, where we think of the coefficients $f_0, f_1, \ldots, f_d$ as formal variables to be determined. Now, we note that for every $i \in \N$, the polynomial $x^i\cdot f^{(i)}$ equals $\sum_{u = i}^{d} x^u \cdot f_u \cdot u(u-1)\cdots (u-i+1) = \sum_{u = i}^d \frac{u!}{(u-i)!}\cdot x^u\cdot f_u$. In particular, the coefficient $f_u$ always appears with the monomial $x^u$ in $x^i\cdot f^{(i)}$ if $u \geq i$, and is zero otherwise. Thus, we have that 
\begin{align*}
Q^{\dagger}_n(x, \partial f) &= \sum_i  \left(\sum_{j = i}^m \frac{n!}{(n+i-j)!}\binom{j}{i}x^{m-j}Q_j(x) \right)x^if^{(i)} \\
&= \sum_i  \left(\left(\sum_{j = i}^m \frac{n!}{(n+i-j)!}\binom{j}{i}x^{m-j}Q_j(x) \right) \cdot \left(\sum_{u = i}^d \frac{u!}{(u-i)!}\cdot x^u\cdot f_u \right)\right) \, .
\end{align*}
Based on the expression above, we note that every coefficient of $Q^{\dagger}_n(x, \partial f)$ is a linear function in the coefficients of $f$. Moreover, every occurrence of $f_j$ in the above expression is in the form $x^jf_j \cdot U(x)$ for some polynomial $U(x)$. Therefore, we get that for every $i$, the coefficient of $x^i$ in $Q^{\dagger}_n(x, \partial f)$ only depends upon $f_0, f_1, \ldots, f_i$ and does not depend on $f_j$ for $j > i$.  
Thus, for any $i \leq k$, seeking an $f$ such that $B(x) + Q^{\dagger}_n(x, \partial f) \equiv 0 \mod x^k$, gives us $k$ linear constraints on $f_0, f_1, \ldots, f_d$, and moreover the $j^{th}$ such constraint (that equates the coefficient of $x^{j-1}$ to zero) only depends on $f_0, f_1, \ldots, f_{j-1}$. Hence, this linear system is lower triangular. We now argue that the diagonal entries of this system are all non-zero. 
 
To get a sense of the diagonal entries of the constraint matrix, we note that the diagonal element in $t^{th}$ row is equal to the coefficient of $f_t$ in the linear form that corresponds to the coefficient of $x^t$ in $Q^{\dagger}_n(x, \partial f)$. We note from the expression for $Q^{\dagger}_n$ that this entry precisely equals 
\[
\sum_{i = 0}^{\max(t, m)} \left(\frac{n!}{(n+i-m)!}\binom{m}{i}\frac{t!}{(t-i)!}Q_m(0)\right) = Q_m(0) \left(\sum_{i = 0}^{\max(t, m)} \frac{n!}{(n+i-m)!}\binom{m}{i}\frac{t!}{(t-i)!}\right) \, .
\]
By expanding the binomial coefficients, and reorganizing, we get 
\begin{align*}
\frac{n!}{(n+i-m)!}\binom{m}{i}\frac{t!}{(t-i)!} &= \frac{n!}{(n-(m-i))!}\cdot \frac{m!}{i!(m-i)!}\cdot \frac{t!}{(t-i)!}\\
&= \frac{n!}{(n-(m-i))!(m-i)!}\cdot m! \cdot \frac{t!}{i!(t-i)!} \\
&= m!\cdot \binom{n}{m-i} \cdot \binom{t}{i}\,. 
\end{align*}
Now, summing over $i$ from $i = 0$ to $\max(m,t)$, we get 
\begin{align*}
\sum_{i = 0}^{\max(m, t)} m!\cdot \binom{n}{m-i} \cdot \binom{t}{i} = m! \binom{n+t}{m} \, ,
\end{align*}
as the sum $\sum_{i = 0}^{\max(m, t)}\binom{n}{m-i} \cdot \binom{t}{i}$ is precisely counting the number of ways of choosing $m$ distinct elements from a collection of $n+t$ distinct elements, which equals $\binom{n+t}{m}$. The diagonal element of the $(t+1)^{st}$ row of the constraint matrix equals $m! \binom{n+t}{m} Q_m(0)$, which is non-zero since $Q_m(0)$ is non-zero, and the characteristic of the the field is either zero or exceeds $\max(n+k, m)$. Finally, we note that the coefficients of $x^0, x, x^2, \ldots, x^{k-1}$ in $Q^{\dagger}_n(x, \partial f)$ only depend on the coefficients $f_0, f_1, \ldots, f_{k-1}$ of $f$, and thus the constraint matrix is actually $k \times k$. Thus, the solution to the equation $B + Q^{\dagger}_n(x, \partial f) \equiv 0 \mod x^k$ is unique modulo $x^k$.

The above proof for the first item in the conclusion of the lemma also sheds light on the proof of the second part. For any solution $f$ of $B + Q^{\dagger}_n(x, \partial f) \equiv 0 \mod x^k$, and $\ell < k$, let us decompose $f$ as $f = h + x^{\ell}\cdot g$, where $h$ consists of monomials of $f$ of degree less than $\ell$ and $x^{\ell}g$ consists of monomials of $f$ of degree at least $\ell$. From the structure of the linear system discussed in the proof of uniqueness above, we immediately get that $h$ must be the \emph{unique} solution of $B + Q^{\dagger}_n(x, \partial h) \equiv 0 \mod x^{\ell}$. Now, by the linearity of the derivative operator and that of $Q$, we have 
\begin{align*}
B + Q(x, \partial f) &= B + Q^{\dagger}_n(x, \partial h + \partial (x^{\ell}g)) \, \\
&= B + Q(x, \partial h) + Q^{\dagger}_n(x, \partial (x^{\ell}g))\,.
\end{align*}
Since $h$ is a solution for  $B + Q^{\dagger}_n(x, \partial h) \equiv 0 \mod x^{\ell}$, we have that there is a polynomial $\tilde{B}(x)$ such that 
\[
x^{\ell}\tilde{B} = B + Q^{\dagger}_n(x, \partial h) \, .
\]
Thus, if $f$ is a solution of $B + Q^{\dagger}_n(x, \partial f) \equiv 0 \mod x^k$, we get that the function $f$ satisfies $x^{\ell}\tilde{B} + Q^{\dagger}_n(x, \partial (x^{\ell}g)) \equiv 0 \mod x^{k}$. To complete the proof of the lemma, we now need to deduce that $g$ satisfies $\tilde{B} + Q^{\dagger}_{n+\ell}(x, \partial g) \equiv 0 \mod x^{k-\ell}$, which we do now. 

\begin{align*}
Q^{\dagger}_n(x, \partial (x^{\ell}g)) &= \sum_{i = 0}^m \left(\sum_{j = i}^m \frac{n!}{(n+i-j)!}\binom{j}{i}x^{m-j}Q_j(x) \right)x^i(x^{\ell}g)^{(i)} \\
&= \sum_{i = 0}^m \left(\sum_{j = i}^m \frac{n!}{(n+i-j)!}\binom{j}{i}x^{m-j}Q_j(x) \right)x^i\left( \sum_{u = 0}^i \binom{i}{u} (x^{\ell})^{(i-u)} g^{(u)}\right) \\
&= \sum_{u = 0}^m g^{(u)} \left(\sum_{i = u}^m \binom{i}{u} x^i (x^{\ell})^{(i-u)} \left(\sum_{j = i}^m \frac{n!}{(n+i-j)!}\binom{j}{i}x^{m-j}Q_j(x) \right) \right)\\
&= \sum_{u = 0}^m g^{(u)} \left(\sum_{i = u}^m \sum_{j = i}^m \frac{n!}{(n+i-j)!} \binom{i}{u}\binom{j}{i} x^{m - j + i} (x^{\ell})^{(i-u)} Q_j(x) \right) \\
&= \sum_{u = 0}^m g^{(u)} \left(\sum_{i = u}^m \sum_{j = i}^m \frac{n!}{(n+i-j)!} \binom{i}{u}\binom{j}{i} x^{m - j + u + \ell} \frac{\ell !}{(\ell - (i - u))!} Q_j(x) \right) \\
&= x^{\ell} \left( \sum_{u = 0}^m x^u g^{(u)} \left(\sum_{i = u}^m \sum_{j = i}^m \frac{n!}{(n+i-j)!} \binom{i}{u}\binom{j}{i} x^{m - j} \frac{\ell !}{(\ell - (i - u))!} Q_j(x) \right) \right)\\
&= x^{\ell} \left( \sum_{u = 0}^m x^u g^{(u)} \left(\sum_{i = u}^m \sum_{j = i}^m \binom{\ell}{i - u} \binom{n}{j-i} \frac{j!}{u!}  x^{m - j}  Q_j(x) \right) \right)\\
&= x^{\ell} \left( \sum_{u = 0}^m x^u g^{(u)} \left(\sum_{j = u}^m x^{m-j} Q_j(x)\cdot  \frac{j!}{u!}\left(\sum_{i = u}^j\binom{\ell}{i - u} \binom{n}{j-i} \right) \right)\right)\, .
\end{align*}
Now, simplifying this further by using $\sum_{i = u}^j\binom{\ell}{i - u} \binom{n}{j-i} = \binom{n + \ell}{j-u}$, we get
\begin{align*}
Q^{\dagger}_n(x, \partial (x^{\ell}g)) &= x^{\ell} \left( \sum_{u = 0}^m x^u g^{(u)} \left(\sum_{j = u}^m x^{m-j} Q_j(x)\cdot  \frac{j!}{u!} \binom{n + \ell}{j-u}\right)\right)\\
&=x^{\ell} \left( \sum_{u = 0}^m x^u g^{(u)} \left(\sum_{j = u}^m x^{m-j} Q_j(x) \cdot  \frac{(n+\ell)!}{(n + \ell - j + u)!} \binom{j}{u}\right)\right)\,.
\end{align*}
Rearranging and re-indexing by using $i$ for $u$, we get that 
\begin{align*}
&x^{\ell}\tilde{B} + Q^{\dagger}_n(x, \partial (x^{\ell}g)) \equiv 0 \mod x^k \\
&\iff x^{\ell}\left(\tilde{B} + \sum_{i = 0}^m x^ig^{(i)}\left(\sum_{j = i}^m \frac{(n + \ell)!}{(n + \ell + i - j)!} \binom{j}{i} x^{m-j}Q_j\right)\right) \equiv 0 \mod x^k\,. \\
\end{align*}
Clearly, this happens if and only if 
\[
\left(\tilde{B} + \sum_{i = 0}^m x^ig^{(i)}\left(\sum_{j = i}^m \frac{(n + \ell)!}{(n + \ell + i - j)!} \binom{j}{i} x^{m-j}Q_j\right)\right) \equiv 0 \mod x^{k-\ell} \, 
\]
or, equivalently, $g$ is the solution for $\tilde{B} + Q^{\dagger}_{n + \ell}(x, \partial g) \equiv 0 \mod x^{k-\ell}$, which is what we set out to prove. This completes the proof of the lemma.  
\end{proof}

\begin{proof}[Proof of \autoref{lem:solution-space-original}]
Like the proof of the second part of \autoref{lem:solution-space-dagger}, the proof of this lemma again involves relying on the structure of the differential equations and some arithmetic based on properties of binomial coefficients. We start with an arbitrary $h$ of degree less than $m$ and seek to find a $g$ such that  $f := h + x^mg$ is a solution of the equation $Q(x, \partial f) \equiv 0 \mod x^k$. Moreover, we need to argue that this $g$ must have the properties asserted in the lemma.  

By the linearity of $Q$ in $\vecy$ variables, and the linearity of the derivative operator, we  have that for any $f$ of the form $h + x^mg$, 
\[
Q(x, \partial f) = Q(x, \partial (h + x^mg)) = Q(x, \partial h) + \sum_{i=0}^m Q_i(x)\cdot (x^mg)^{(i)} \, .
\]
Now, we reorganise the expression $Q(x, x^mg)$ a bit in a more usable form. 
\begin{align*}
\sum_{i=0}^m Q_i(x)(x^mg)^{(i)} &= \sum_{i=0}^m Q_i(x)\cdot \left(\sum_{j=0}^i\binom{i}{j}(x^m)^{(i-j)}g^{(j)}\right)\\
&= \sum_{j=0}^m \left(\sum_{i=j}^{m}Q_i(x)\cdot \binom{i}{j}(x^m)^{(i-j)}\right)g^{(j)}\\
&= \sum_{j=0}^m \left(\sum_{i=j}^{m}Q_i(x)\cdot \binom{i}{j} \frac{m!}{(m-i + j)!}x^{m-i+j}\right)g^{(j)}\\
&= \sum_{j=0}^m x^j\left(\sum_{i=j}^{m}\binom{i}{j} \frac{m!}{(m-i + j)!}x^{m-i}Q_i\right)g^{(j)}\\
&= Q^{\dagger}_m(x, \partial g) \,.
\end{align*}
From \autoref{lem:solution-space-dagger}, we know that for every polynomial $B$, and natural numbers $m$, $k$ if the underlying field has characteristic zero or larger than $m+k$, then the equation $B(x) + Q^{\dagger}_m(x, \partial g) \equiv 0 \mod x^k$ has a unique solution. So, if we set $B = Q(x, \partial h)$, and consider the unique solution $g$ of $B(x) + Q^{\dagger}_m(x, \partial g) \equiv 0 \mod x^k$, we have that $Q(x, h + x^mg) = Q(x, \partial h) + Q^{\dagger}_m(x, \partial g)$, which is zero modulo $x^k$ by the choice of $g$. This completes the proof of the lemma. 
\end{proof}

\section{Pruning the list to constant size}

 In this section, we recall a beautiful recent result of Kopparty, Ron-Zewi, Saraf and Wootters \cite{KoppartyRSW2023} and a subsequent improvement to it due to Tamo \cite{Tamo2024}, who showed that for any constant $\epsilon > 0$, univariate multiplicity codes (and folded Reed-Solomon codes) can be list decoded with fractional agreement $(1-\delta + \epsilon)$ with constant list size in polynomial time, where $\delta$ is the relative distance of the code. One of their main technical insights is a property of these codes, which essentially shows that any low-dimensional affine space cannot contain \emph{too many} codewords that are \emph{close} to a received word. Moreover, they give a randomized algorithm that given a low-dimensional subspace via its basis, and a received word, output this constant-size list in polynomial time. We use this algorithm  off-the-shelf for our algorithm. We describe the algorithm, and state the main technical theorem (or rather, a special case for univariate multiplicity codes) that we use, and refer the interested reader to \cite{Tamo2024, KoppartyRSW2023} for the general statement and a proof.  
 
\newcommand{\prune}{\texttt{Prune}\xspace}
\noindent \begin{minipage}{\algwidth}
\begin{algorithm}[H]
\caption{\prune (Reducing to constant list size \cite{KoppartyRSW2023})} 
\label{algo:prune}
\KwIn{
  Codewords $c_0, c_1, \ldots, c_{m}$ and received word $R = (\alpha_j, \beta_{j, 0}, \ldots, \beta_{j, s-1})_{j=1}^{n}$ for univariate multiplicity codes of degree $d$, block length $n$ and multiplicity parameter $s$ and  natural numbers $t, k$.
}
\KwOut{
  A list ${\cal L}$ of codewords in the affine span of $c_0, c_1, \ldots, c_{m}$.
}
\nonl\hrulefill\\
Set $\cal L \gets \emptyset$ (empty set) ;\\
\For{$j\gets 1$ to $k$}
{
  Pick $i_1, i_2, \ldots, i_t \in \{1, 2, \ldots n\}$ independently and uniformly at random ;\\
  \lIf{there is exactly one codeword $\tilde{c}$ in the affine span of $c_0, c_1, \ldots, c_{m}$ that agrees with $R$ on coordinates $i_1, i_2, \ldots, i_t$}{add $\tilde{c}$ to $\cal L$}
}
\KwRet{$\cal L$ .}
\end{algorithm}
\end{minipage}

The main theorem about the algorithm that we use in our proof is the following. The statement below is a quantitative strengthening of the statement of Kopparty, Ron-Zewi, Saraf and Wootters due to Tamo.  
\begin{theorem}[{\cite[Lemma 3.1]{Tamo2024}}]\label{thm:prune-krsw}
Let $\epsilon >0 $ be an arbitrary constant, $\cal C$ be the affine span of independent codewords $c_0, c_1, \ldots, c_{m}$ of univariate multiplicity codes of degree $d$, block length $n$ and multiplicity $s$. Let $\delta = 1-d/sn$  be the fractional distance of the code. 

If  \[t = m   \quad \text{and} \quad  k =  O\left(\frac{ \log \left(1/\epsilon\right)^m}{\epsilon^m} \right) \]
then, with with high probability, the list $\cal L$  output by \prune (\autoref{algo:prune}) on input $c_0, \ldots, c_{m}$, and $R, k, t$ includes all codewords of $\cal C$ that have fractional agreement at least $(1-\delta + \epsilon)$  with the received word $R$. 

Moreover, the algorithm requires at most $\poly(mtk)$ field operations in the worst case.  
\end{theorem}
We refer to \cite{Tamo2024, KoppartyRSW2023} for a proof of the first part of the theorem. The bound on the time complexity follows from the fact that the algorithm essentially solves $k$ instances of a linear system with $m+1$ variables and $t$ constraints each. 
 
\section{Fast decoding the multiplicity code \texorpdfstring{({\cref{thm:intro-main-large-multiplicity,thm:main-large-multiplicity}})}{}}\label{sec:mult-list-decoding}
We are now ready to prove the main theorems for fast list decoding of multiplicity codes.

\begin{proof}[Proof of \autoref{thm:main-large-multiplicity}]
Let the parameter $s_0$ in the theorem be set to $\lceil (1 + 2/\epsilon)^2 \rceil$, and let $s$ be any integer greater than $s_0$. Let $m$ be set to $\lceil 2/\epsilon \rceil$, which, by the choice of $s$ satisfies that $s/m - 1 \geq 2/\epsilon$. 

Given the received word $R$ the decoder uses \autoref{thm:multivariate-interpolation} to construct a non-zero polynomial $Q(x, \vecy)$ \emph{explaining} the received word, with its $x$-degree being at most $D \leq n(s-m)/m$ and $\vecy$ degree $1$. This takes time at most $\tilde{O}(n\poly(s))$. 

Now, by \autoref{lem:close-enough-codewords-satisfy-the-differential-equation}, we have for the above $Q$ and any message polynomial $f$ whose encoding agrees with the received word $R$ on at least $(D+d)/(s-m)$ coordinates, the polynomial $Q(x, \partial f)$ is identically zero. 

From the discussion at the beginning of \autoref{sec:solving-differential-equations}, we have that $Q_m(0)$ can be assumed to be non-zero without loss of generality. 

From \autoref{thm:fast-differential-equation-solver}, we have that the space of degree $d$ solutions of $Q(x, \partial f)$ is an affine space of dimension $m$, and \solveQ (\autoref{algo:solve-original}) outputs a basis for this affine space using at most $\tilde{O}((D + d)\poly(m)) \leq \tilde{O}(n + d)\poly(s)$ many field operations. This concludes the proof of \autoref{thm:main-large-multiplicity}.\end{proof}

\begin{proof}[Proof of \autoref{thm:intro-main-large-multiplicity}]
Finally, \autoref{thm:main-large-multiplicity} using \autoref{thm:prune-krsw}, we have that a  list of codewords consisting of all the true solutions can be output in time $O\left(n(sn/d)^{\poly(1/\epsilon)}\right)$ with probability at least $0.99$. This includes the time to compute the parameters $t, k$ required by \prune (\autoref{algo:prune}) as inputs. The size of the list of codewords is at most $O\left((d/sn)^{\poly(1/\epsilon)}\right)$. 
  
Moreover, we have that the above algorithm works as long as the agreement between the received word and the codewords is larger than 
\begin{align*}
(D + d)/(s-m) &\leq d/(s-m) + D/(s-m) \\
&\leq d/(s-m) + n/m\\
&\leq d/s + (dm)/(s(s-m)) + n/m
\end{align*} 
Thus the fractional agreement needed is at least $d/sn + (d/n)(m/(s(s-m))) + 1/m$. By the choice of parameters, this is at most $d/sn + \epsilon$, as stated in the theorem. 
\end{proof}


\section{Fast interpolation for Folded Reed-Solomon codes}\label{sec:frs-constructing-functional-equations}
We start with the following definition that the natural analogue of \autoref{defn:differential-operator-for-derivative} in the context of Folded Reed-Solomon codes and was defined in this form in \cite{GuruswamiW2013}. $\gamma$ is a field element of sufficiently high order throughout this section.  

\begin{definition}[Iterated folding operator]\label{defn:folding-operator}
Let $s \in \N$ be a parameter. Then, for every $m$ such that $0 \leq m < s$, the function $\psi$ is an $\F$-linear map from the set of polynomials in $\F[x, y_0, y_1, \ldots, y_m]$ with $\vecy$ degree $1$ to the set of polynomials in $\F[x, y_0, y_1, \ldots, y_{m+1}]$ with $\vecy$ degree $1$ and is defined as follows. 
\[
\psi\left(\tilde{Q}(x) + \sum_{i = 0}^m Q_i(x)\cdot y_i\right) := \tilde{Q}(\gamma x) + \sum_{i = 0}^m Q_i(\gamma x)\cdot y_{i+1} \, .
\]

For an integer $i \leq (s-m) $, we use $\psi^{(i)}$ to denote the linear map from $\F[x, y_0, \ldots, y_m]$ to $\F[x, y_0, \ldots, y_{m + i}]$ obtained by applying the operator $\psi$ iteratively $i$ times.
\end{definition}

We now state the main theorem of this section. 
\begin{theorem}\label{thm:frs-multivariate-interpolation}
Let $R = (\alpha_i, \beta_{i, 0}, \ldots, \beta_{i, s-1})_{i = 1}^n$ be a received word for a FRS code decoder, and let $m \leq s$ be any parameter. 

Then, for $D \leq ((n(s-m))/m)$, there exists a non-zero polynomial $Q(x, y_0, y_1, \ldots, y_m)$ of the form $Q = \tilde{Q}(x) + \sum_i Q_i(x) y_i$, with degree of $\tilde{Q}$ and each $Q_i$ being at most $D$ such that for all $i \in \{0, 1, \ldots, s-m-1\}$,
\[
\psi^{(i)}(Q)(\alpha_j, \beta_{j, 0}, \ldots, \beta_{j, s-1}) = 0 \ \forall j \in \{ 1, 2, \dotsc, n \} .
\]

Moreover, there is a deterministic algorithm that when given $R$ as input, runs in time $\tilde{O}(n\poly(sm))$ and outputs such a polynomial $Q$ as a list of coefficients. 
\end{theorem}

\subsection{Proof of \autoref{thm:frs-multivariate-interpolation}}
The proof of \autoref{thm:frs-multivariate-interpolation} is exactly on the same lines as that of \autoref{thm:multivariate-interpolation}, upto some variation in the details. We start with the definition of an appropriate lattice, whose shortest vector would ultimately give us the polynomial $Q$ that we are looking for. Owing to the similarities to the proofs in the algorithm for multiplicity codes, we try to keep this presentation short and succinct, and take the liberty to skip a few details.


We start by defining the lattice. 
\begin{definition}\label{def:frs-lattice-definition}
Let $R = (\alpha_i, \beta_{i, 0}, \ldots, \beta_{i, s-1})_{i = 1}^n$ be a received word for an FRS code decoder, and let $m \leq s$ be any parameter. 

For $i \in \{0, 1, \ldots, m\}$, let $A_i(z)$ be a univariate polynomial of degree at most $n(s+1-i)$ in $\F[z]$ such that for every $j \in \{1, 2, \ldots, n\}$ and $k \in \{0, ,1, \ldots, s - m-1\}$, 
\[
A_i(\gamma^k\alpha_j) = \beta_{j, i + k} \, .
\]
Let $\mathcal{L}_R^{(m)}$ be the $(m+2)$-dimensional lattice over the ring $\F[x]$ defined as 
\[
\mathcal{L}_R^{(m)} := \left\{\tilde{h}(x) \cdot \prod_{i = 1}^n \prod_{j = 0}^{s-m-1}(x - \gamma^j\alpha_{i}) + \sum_{i = 0}^m h_i(x)\cdot (y_i - A_i(x)) \colon \tilde{h}, h_i \in \F[x] \right\}\,.\qedhere
\]
\end{definition} 
The following lemma asserts that the polynomials $A_i's$ can be computed in nearly-linear time, via an off-the-shelf application of fast polynomial interpolation, e.g., \cite[Algorithm 10.11]{GathenG-MCA}.  
\begin{theorem}[{\cite[Algorithm 10.11]{GathenG-MCA}}]\label{thm:fast-polynomial-interpolation}
Let $m \in \N$ be a parameter with $m \leq s$. Then, there is a deterministic algorithm that when given the received word $R = (\alpha_i, \beta_{i, 0}, \ldots, \beta_{i, s-1})_{i = 1}^n$ as input, outputs the polynomials $A_0, A_1, \ldots, A_m$ defined in \autoref{def:frs-lattice-definition} in time $\tilde{O}(nsm)$. 
\end{theorem}


The following lemma summarizes some of the crucial properties of the lattice defined in \autoref{def:frs-lattice-definition}.
\begin{lemma}\label{lem:frs-properties-of-lattice}
Let $R = (\alpha_i, \beta_{i, 0}, \ldots, \beta_{i, s-1})_{i = 1}^n$ be a received word for a FRS code decoder, and let $m \leq s$ be any parameter, and let $\mathcal{L}_R^{(m)}$ be the lattice as defined in \autoref{def:frs-lattice-definition}. Then, the following are true. 
\begin{itemize}
\item Any non-zero polynomial $Q$ in $\mathcal{L}_R^{(m)}$ is of the form $Q(x, y_0, y_1, \ldots, y_m) = \tilde{Q}(x) + \sum_{\ell} Q_{\ell}(x) \cdot y_{\ell}$ and for 
all $i \in \{0, 1, \ldots, s-m-1\}$,
\[
\psi^{(i)}(Q)(\alpha_j, \beta_{j, 0}, \ldots, \beta_{j, s-1}) = 0 \ \forall j \in \{ 1, 2, \dotsc, n \}\, .
\]
\item There is a non-zero polynomial in $\mathcal{L}_R^{(m)}$ with $x$-degree at most $D \leq (n(s-m)/m)$. 
\item Such a polynomial can be found in time $\tilde{O}(n\poly(sm))$.
\end{itemize}

\end{lemma}
\begin{proof}[Proof sketch]
The proof follows the proof of \autoref{lem:properties-of-lattice} in general, and in particular, the proofs of the second and the third item are exactly the same.  For the first item, we briefly sketch some of the details.

Let $Q$ be a nonzero polynomial in $\mathcal{L}_R^{(m)}$. Thus, there are polynomials $\tilde{h}$ and $h_0, h_1, \ldots, h_m$ in $\F[x]$ such that 
\[
Q = \tilde{h}(x) \cdot \prod_{\ell = 1}^n \prod_{j = 0}^{s-m-1} (x - \gamma^j\alpha_{\ell}) + \sum_{\ell = 0}^m h_{\ell}(x)\cdot (y_{\ell} - A_{\ell}(x)) \, .
\]
By rearranging and separating out the terms depending on the $\vecy$ variables, we get the desired form. Now, just from its definition, it is quite clear that for every $i \in \{0, 1, \ldots, s-m-1\}$, and for every $\ell \in \{1, 2, \ldots, n\}$, the polynomial $\prod_{\ell = 1}^n \prod_{j = 0}^{s-m-1} (x - \gamma^j\alpha_{\ell})$ vanishes on $\gamma^i\alpha_{\ell}$, and thus, $\psi^{(i)}\left(\prod_{\ell = 1}^n \prod_{j = 0}^{s-m-1} (x - \gamma^j\alpha_{\ell})\right)$ vanishes on input $\alpha_{\ell}$.

The following claim, whose proof is immediate from the definition of the operator $\psi$ and the polynomials $A_i$, together with the linearity of the operator $\psi$ then completes the proof of the first item. 
\end{proof}



\begin{claim}\label{clm:frs-yi-Ai-under-psi}
For every $i \in \{0, 1, \ldots, s-m-1\}$, $j \in \{1, 2, \ldots, n\}$ and $\ell \in \{0, 1, \ldots, m\}$, we have that the polynomial $\psi^{(i)}(h_{\ell}(x)(y_{\ell} - A_{\ell}(x)))$ when evaluated at $(\alpha_j, \beta_{j, 0}, \ldots, \beta_{j, s})$ is zero.
\end{claim}
\begin{proof}[Proof sketch of \autoref{thm:frs-multivariate-interpolation}]
The proof of \autoref{thm:frs-multivariate-interpolation} is an immediate consequence of \autoref{def:frs-lattice-definition}, \autoref{thm:fast-polynomial-interpolation} and \autoref{lem:frs-properties-of-lattice}. We skip the details. 
%
\end{proof}

\subsection{Close enough codewords satisfy the equation}
We end this section with the following lemma, which is an analogue of \autoref{lem:close-enough-codewords-satisfy-the-differential-equation} and notes all message polynomials whose encoding is close enough to the received word satisfy a natural functional equation. Thus, solving this equation fast is sufficient for FRS list decoding. 

\begin{lemma}\label{lem:frs-close-enough-codewords-satisfy-the-differential-equation}
Let $R = (\alpha_i, \beta_{i, 0}, \ldots, \beta_{i, s-1})_{i = 1}^n$ be a received word for a multiplicity code decoder, $m \leq s$ be a parameter, and let $Q(x, y_0, y_1, \ldots, y_m) \in \mathcal{L}_R^{(m)}$ be a non-zero polynomial with $x$-degree at most $D \leq (n(s-m)/m)$.  

Then, for every polynomial $f \in \F[x]$ of degree at most $d$ such that its FRS encoding (with folding parameter $s$) agrees with $R$ on more than $(D + d)/(s-m)$ values of $i$, $Q(x, f, f(\gamma x), \ldots, f(\gamma^mx))$ is identically zero. 

\end{lemma}

\section{Solving the functional equation in nearly-linear time}\label{sec:frs-solving-functional-equations}
In this section, we design faster algorithms for \emph{solving} functional equations of the form \[Q\left(x, f(x), \ldots, f(\gamma^{m}x)\right) \equiv 0\] obtained in \autoref{lem:frs-close-enough-codewords-satisfy-the-differential-equation}. Since the degree of $Q$ in $\vecy$ variables is at most $1$, we have that the solution space of all polynomials of degree at most $d$ satisfying the above equation is an affine space. The following theorem of  Guruswami \& Wang \cite{GuruswamiW2013} bounds the dimension of this space by $m$ and outputs a basis efficiently. The main content of the theorem below is that this basis can be obtained quite fast. 
\begin{theorem}\label{thm:frs-fast-functional-equation-solver-original}
Let $\F$ be any field, let $\gamma \in \F$ be an element of order greater than $d$, and let \[Q(x, y_0, \ldots, y_m) = \tilde{Q}(x) + \sum_{i = 0}^m Q_i(x)\cdot y_i\] be a non-zero polynomial with $x$-degree at most $D$. Then, the affine space of polynomials $f(x) \in \F[x]$ of degree at most $d$ that satisfy 
\[
Q\left(x, f(x), f(\gamma x), \ldots, f(\gamma^{m} x)\right) \equiv 0
\]
has dimension at most $m$.

  Moreover, there is a deterministic algorithm that when given $Q$ as an input via its coefficient vector, and the parameter $d$, performs at most $\tilde{O}((D+d)\poly(m))$ field operations and outputs a basis for this affine space. 
\end{theorem}  
Before proceeding further, we set up some notations and conventions that will simplify our presentation. 

For any polynomial $Q(x, \vecy) = \tilde{Q}(x) + \sum_{i = 0}^m Q_i(x)\cdot y_i$ that is non-zero, we note that we can assume without loss of generality that there exists an $i$ such that the constant term of $Q_i$, or in other words, $Q_i(0)$ is non-zero. If this is not the case, that means that for every $i$, $Q_i$ must be divisible by $x$, and thus, for a solution of the equation $Q(x, f(x), f(\gamma x), \ldots, f(\gamma^{m} x))     \equiv 0$ to exist, it must be the case that $\tilde{Q}$ is also divisible by $x$. Thus, in this case, we can instead just work with the polynomial $Q/x$. Peeling out common factors of $x$ in this way eventually leads us to the case that at least one of the $Q_i$s has a non-zero constant term. Moreover, this procedure can be done in linear time in the length of the coefficient vector of $Q$. Thus, going forward, we assume that there is at least one $i$ such that $Q_i(0)$ is non-zero. 

More generally, let $B_Q \subseteq \{0, 1, \ldots, m\}$ be the (non-empty) subset of all indices $i$ such that $Q_i(0)$ is non-zero. Moreover, for a given polynomial $Q(x, \vecy) = \tilde{Q}(x) + \sum_{i = 0}^m Q_i(x)\cdot y_i$, we denote by $P_Q(z)$ the univariate polynomial defined as follows. 
\[
P_Q(z) := \sum_{i \in B_Q} Q_i(0)z^i \, .
\]
From everything discussed in the previous paragraph, we have that $P_Q$ is a non-zero zero polynomial of degree $\max{B_Q} \leq m$. Thus, it has at most $m$ zeros in the field $\F$ and in particular, in the subset $\{1, \gamma, \gamma^2, \ldots, \gamma^d\}$ of the field. Let $Z_{Q, k}$ denote the set 
\[
Z_{Q,k} := \{i \colon  i \in \{0, 1, \ldots, k-1\} \text{ such that } P_Q(\gamma^i) = 0\} \, .
\]
We now state a result of Guruswami \& Wang \cite{GuruswamiW2013} about structure of solutions of the equation $Q(x, f(x), f(\gamma x) \ldots, f(\gamma^{m}x)) \equiv 0$. 
\begin{theorem}[\cite{GuruswamiW2013}]\label{thm:frs-dim-of-solution-space-guruswami-wang}
Let $k$ be any natural number, $\F$ be any field and $\gamma$ be an element of order at least $d$. Let \[Q(x, y_0, \ldots, y_m) = \tilde{Q}(x) + \sum_{i = 0}^m Q_i(x)\cdot y_i\] be such that its $x$-degree is at most $D$ and $Q_i(0)$ is non-zero for some $i \in \{0,1,\dots,m\}$. 
Then, the following are true. 
\begin{itemize}
\item The set of polynomials $f$ of degree at most $(k-1)$ that satisfy \[Q(x, f(x), f(\gamma x) \ldots, f(\gamma^{m}x)) \equiv 0  \mod x^k \, .\] form an affine space of dimension $m$. 
\item There is a $\poly(k, D, m)$ time algorithm that when given the coefficient vector of $Q$ and the parameter $d$ as inputs, outputs a basis for this affine space.
\end{itemize}


\end{theorem}  
The following technical lemma is implicit in the work of Guruswami \& Wang, and essentially drives the proof of their theorem stated above. This also turns out to be crucial to our proof.
\begin{lemma}[Implicit in \cite{GuruswamiW2013}]\label{lem:frs-structure-of-solution-space-guruswami-wang}
Let $k$ be any natural number, $\F$ be any field and $\gamma$ be an element of order at least $d$. Let \[Q(x, y_0, \ldots, y_m) = \tilde{Q}(x) + \sum_{i = 0}^m Q_i(x)\cdot y_i\] be such that its $x$-degree is at most $D$ and $Q_i(0)$ is non-zero for some $i \in \{0,1,\dots,m\}$. Let $P_Q$ and $Z_{Q,k}$ be as defined earlier in this section. Then, the following are true. 
\begin{itemize}
\item \label{item:frs-unique-extension} For every polynomial $h$ of degree less than $k-1$ that is supported only on monomials $x^i$ with $i \in Z_{Q, k}$, there is a unique polynomial $f$ of degree less than $k$ such that $Q(x, f(x), f(\gamma x) \ldots, f(\gamma^{m}x)) \equiv 0  \mod x^k$ and for all $i \in Z_{Q, k}$, the coefficient of $x^i$ in $h$ equals the coefficient of $x^i$ in $f$. 
\item \label{item:frs-basis} For every $i \in Z_{Q, k}$, there is a unique polynomial $g_i(x)$ of degree less than $k$, supported on monomials in the set $\{x^i \colon i \in \{0, 1, \ldots, d\}\setminus Z_{Q, k}\}$ such that the set $\{f_i \colon i \in Z_{Q, k}\}$ of polynomials forms a basis of this affine space of solutions, where $f_i := x^i + g_i$. 
\item There is a $\poly(k, D, m)$ time algorithm that when given the coefficient vector of $Q$ and the parameter $d$ as inputs, outputs $\{f_i \colon i \in Z_{Q, k}\}$.
\end{itemize}
\end{lemma}

Our main result in this section is an algorithm that improves the running time of the algorithm of Guruswami \& Wang in \autoref{thm:frs-dim-of-solution-space-guruswami-wang} from $\poly(k, D, m)$ to $\tilde{O}((D+k)\poly(m))$. As in the fast list decoder for solving differential equations in \autoref{sec:solving-differential-equations}, our algorithm proceeds via divide and conquer. 

The following technical lemma provides the set up for the algorithm based on the divide and conquer paradigm. 

\begin{lemma}\label{lem:frs-divide-conquer}
Let $k$ be any natural number, $\F$ be any field and $\gamma$ be an element of order at least $d$. Let $Q(x, y_0, \ldots, y_m) = \tilde{Q}(x) + \sum_{i = 0}^m Q_i(x)\cdot y_i$ be such that its $x$-degree is at most $D$ and $Q_i(0)$ is non-zero for some $i \in \{0,1,\dots,m\}$. Let $P_Q$ and $Z_{Q, k}$ be as defined earlier in this section. Let $f(x)$ be a polynomial of  degree at most $k-1$ that is a solution of $Q(x, f(x), f(\gamma x) \ldots, f(\gamma^{m}x)) \equiv 0  \mod x^k \, ,$ and let $\ell \in \{0, 1, \ldots, k-1\}$ be any natural number. 

Then, there exist polynomials $h(x), g(x)$ of degree less than $\ell$ and $k-\ell$ respectively, satisfying $f = h(x) + x^{\ell}\cdot g(x)$ such that the following are true. 
\begin{itemize}
\item $h$ is the unique solution of \[Q(x, h(x), h(\gamma x) \ldots, h(\gamma^{m}x)) \equiv 0  \mod x^\ell\] that has the same coefficient as $f$ on all monomials $x^i$, with $i \in \{0, 1, \ldots, \ell-1\} \cap Z_{Q, k}$. 
\item $g$ is the unique solution of  \[
  x^{-\ell}Q\left(x, h(x), h(\gamma x) \ldots, h(\gamma^{m}x)\right) + \sum_{j = 0}^m {\gamma^{j\ell}} Q_j(x) \cdot g(\gamma^{j} x) \equiv 0 \mod x^{k-\ell} 
  \] such that for all $i \in \{\ell, \ell + 1, \ldots, k-1\} \cap Z_{Q, k}$, the coefficient of $x^{i - \ell}$ in $g$ equals the coefficient of $x^i$ in $f$. 
\end{itemize} 
\end{lemma}
\begin{proof}
We set $h$ to be the unique polynomial of degree than $\ell$ that agrees with $f$ on all monomials of degree less than $\ell$ and we define $g$ to be equal to $x^{-\ell}(f-h)$. Since $f$ and $h$ agree on all monomials of degree less than $\ell$, $(f-h)$ is divisible by $x^{\ell}$ and $g$ as defined here is indeed a polynomial and has degree less than $k-\ell$. In the rest of this proof we argue that these $g, h$ have the properties stated in the lemma. 

\paragraph*{Properties of $h$: }
  Since $Q(x, f(x), f(\gamma x) \ldots, f(\gamma^{m}x)) \equiv 0  \mod x^k$, and $\ell < k$, we have that $Q(x, f(x), f(\gamma x) \ldots, f(\gamma^{m}x) )\equiv 0  \mod x^\ell$. Now, substituting $f$ by $h + x^{\ell}g$, we get that $h$ must satisfy $Q(x, h(x), f(\gamma x) \ldots, h(\gamma^{m}x) )\equiv 0  \mod x^\ell$. By definition, $f$ and $h$ have the same coefficient for all monomials of degree less than $\ell$ and in particular for all monomials $x^i$ where  $i \in \{0, 1, \ldots, \ell-1\} \cap Z_{Q, k}$. Furthermore, from the first item in \autoref{item:frs-unique-extension}, we have that once we fix the coefficients for monomials with degree $i$ in $ \{0, 1, \ldots, \ell-1\} \cap Z_{Q, k}$, the solution is unique. This completes the proof of the first item. 
  
  \paragraph*{Properties of $g$: } 
  From $Q(x, f(x), f(\gamma x) \ldots, f(\gamma^{m}x) )\equiv 0  \mod x^k$ and $f = h + x^{\ell}\cdot g$, we get 
  \begin{align*}
  Q\left(x, h(x) + x^{\ell}\cdot g(x), \ldots, h(\gamma^m x) + (\gamma^m x)^{\ell}\cdot g(\gamma^{m} x)\right) &\equiv 0 \mod x^k \\
  \tilde{Q}(x) + \sum_{j = 0}^m Q_j(x) \cdot \left(h(\gamma^j x) + (\gamma^j x)^{\ell}\cdot g(\gamma^{j} x)\right) &\equiv 0 \mod x^k \\
  Q\left(x, h(x), h(\gamma x) \ldots, h(\gamma^{m}x)\right) + \sum_{j = 0}^m Q_j(x) \cdot \left((\gamma^j x)^{\ell}\cdot g(\gamma^{j} x)\right) &\equiv 0 \mod x^k \\
  Q\left(x, h(x), h(\gamma x) \ldots, h(\gamma^{m}x)\right) + x^{\ell}\cdot \left(\sum_{j = 0}^m Q_j(x) \cdot \left((\gamma^j)^{\ell}\cdot g(\gamma^{j} x)\right)\right) &\equiv 0 \mod x^k \\
    Q\left(x, h(x), h(\gamma x) \ldots, h(\gamma^{m}x)\right) + x^{\ell}\cdot \left(\sum_{j = 0}^m {\gamma^{j\ell}} \cdot Q_j(x) \cdot g(\gamma^{j} x)\right) &\equiv 0 \mod x^k \,.
  \end{align*}
  Now, since $h$ satisfies $Q\left(x, h(x), h(\gamma x) \ldots, h(\gamma^{m}x)\right) \equiv 0 \mod x^{\ell}$, we have that the polynomial $Q\left(x, h(x), h(\gamma x) \ldots, h(\gamma^{m}x)\right)$ is divisible by $x^{\ell}$. Thus, $g$ must satisfy 
  \[
  x^{-\ell}\cdot Q\left(x, h(x), h(\gamma x) \ldots, h(\gamma^{m}x)\right) + \sum_{j = 0}^m {\gamma^{j\ell}} \cdot Q_j(x) \cdot g(\gamma^{j} x) \equiv 0 \mod x^{k-\ell} \,.
  \]
  Clearly from its definition, for every $i\in \{\ell, \ell + 1, \ldots, k-1\} \cap Z_{Q, k}$, the coefficient of $x^{i-\ell}$ in $g$ equals the coefficient of $x^i$ in $f$. To complete the proof of the lemma, it suffices to show that there is a unique solution of the above function equation that satisfies this. Let $\hat{Q}(x,\vecy)$ denote the polynomial $x^{-\ell}\cdot Q\left(x, h(x), f(\gamma x) \ldots, h(\gamma^{m}x)\right) + \sum_{j = 0}^m {\gamma^{j\ell}} \cdot Q_j(x)\cdot  y_j$. Then, we have that set $B_Q$ equals the set $B_{\hat{Q}}$. To argue the uniqueness condition, it suffices to show that the set $Z_{\hat{Q}, k-\ell}$ is a subset of $\{ i -\ell\colon i \in \{\ell, \ell + 1, \ldots, k-1\} \cap Z_{Q, k}\}$. From the structure of $\hat{Q}$, we have that 
  \[
  P_{\hat{Q}}(z) = \sum_{j \in B_Q} Q_{j}(0)\gamma^{j\ell} z^j \, . 
  \]
  Thus, for any $i \in \N$, we have 
\[
  P_{\hat{Q}}(\gamma^i) = \sum_{j \in B_Q} Q_{j}(0)\gamma^{(\ell + i)j} \, . 
  \]  
  Thus, $\gamma^i$ is a root of $P_{\hat{Q}}(z)$ if and only if $\gamma^{\ell + i}$ is a root of $P_Q(z)$. Now, since the order of $\gamma$ exceeds $k$, we get that $Z_{\hat Q, k-\ell}$ equals $\{ i -\ell: i \in \{\ell, \ell + 1, \ldots, k-1\} \}\cap Z_{Q, k}$. Therefore, from the first item in  \autoref{lem:frs-structure-of-solution-space-guruswami-wang}, we have that once we fix the coefficients of monomials with degrees in $Z_{\hat{Q}, k-\ell}$, the solution to $\hat{Q}(x, g(x), \ldots, g(\gamma^m x)) \equiv 0 \mod x^{k-\ell}$ is unique, and $g$ as defined above is that unique solution by its definition.    
  
  This completes the proof of the lemma. 
\end{proof}

We are now ready to describe and analyse our algorithm. We first describe the recursive subroutine that we use and then the main algorithm that invod.

\subsection{Algorithm for solving \texorpdfstring{$Q(x, f(x), \ldots, f(\gamma^m x)) \equiv 0$}{Q(x,f(x),..)=0}}

\newcommand{\funcRSolveQ}{\texttt{FuncRSolveQ}\xspace}
\noindent \begin{minipage}{\algwidth}
  \begin{algorithm}[H]
  \caption{\funcRSolveQ (Solving $Q(x, f(x), \ldots, f(\gamma^m x)) \equiv 0 $ recursively)}\label{algo:frs-recurse}
  \KwIn{
    $(Q(x, \vecy), k, t)$, where $Q(x, \vecy) \in \F[x, \vecy]$ is polynomials with $Q(x, \vecy) = \tilde{Q}(x) + \sum_{i = 0}^m Q_i(x)\cdot y_i$ with $Q_m(0) \neq 0$, $x$-degree $D$, natural numbers $k, t$ with either $t \in \{0, 1, \ldots, k-1\} \cap Z_{Q, k}$ or $t = -1$. $\gamma \in \F$ is an element of order at least $k$.}
  \KwOut{
    The unique polynomial $f(x) \in \F[x]$ of degree less than $k$ supported on the set $\{x^i : i \in \{0, 1, \ldots, k-1\} \setminus Z_{Q, k}\} \cup \{x^t\}$ of monomials such that the coefficient of $x^t$ in $f$ is one if $t \geq  0$ and zero if $t < 0$ and $f$ satisfies $Q(x, f(x), \ldots, f(\gamma^m x)) \equiv 0 \mod x^k$.}
  \nonl\hrulefill\\
  Set $\ell \gets \lfloor k/2 \rfloor$ ;\\
  \eIf{$k = 1$}{
    \eIf{$0 \notin Z_{Q,1}$}{
      \KwRet{$-\tilde{Q}(0)/{\left(\sum_{i \in B_{Q}}Q_i(0)\right)}$ ;}
    }
    {
      \leIf{$ t \neq 0$}{\KwRet{$0$}}{\KwRet{$1$}}
    }
  }
  {
    \eIf{$t < \ell$}{
      Run \funcRSolveQ (\autoref{algo:frs-recurse}) on inputs $(Q, \ell, t)$ to obtain the polynomial $h(x)$;\\
      Run \funcRSolveQ (\autoref{algo:frs-recurse}) on inputs $\left(\hat{Q}, k-\ell, -1\right)$ to obtain the polynomial $g(x)$ where\\
      \nonl \Indp \Indp $\hat{Q}(x,\vecy):= x^{-\ell}\cdot Q\left(x, h(x), h(\gamma x) \ldots, h(\gamma^{m}x)\right) + \sum_{j = 0}^m {\gamma^{j\ell}} \cdot Q_j(x) \cdot y_j\,;$
    }
    {
      Run \funcRSolveQ (\autoref{algo:frs-recurse}) on inputs $(Q, \ell, -1)$ to obtain polynomial $h(x)$;\\
      Run \funcRSolveQ (\autoref{algo:frs-recurse}) on inputs $\left(\hat{Q}, k-\ell, t-\ell \right)$ to obtain the polynomial $g(x)$ where\\
      \nonl \Indp \Indp $\hat{Q}(x,\vecy) := x^{-\ell}\cdot Q\left(x, h(x), h(\gamma x) \ldots, h(\gamma^{m}x)\right) + \sum_{j = 0}^m {\gamma^{j\ell}} \cdot Q_j(x) \cdot y_j\,;$
    }
    \KwRet{$h(x) + x^{\ell}\cdot g(x)$.}
  }
  \end{algorithm} 
  \end{minipage}

\newcommand{\funcSolveQ}{\texttt{FuncSolveQ}\xspace}
\noindent \begin{minipage}{\algwidth}
  \begin{algorithm}[H]
  \caption{\funcSolveQ (Solving $Q(x, f(x), \ldots, f(\gamma^m x)) \equiv 0$)} \label{algo:frs-solve-original}
  \KwIn{
    $(Q(x, \vecy), k)$, where $Q(x, \vecy) \in \F[x, \vecy]$ is polynomials with $Q(x, \vecy) = \tilde{Q}(x) + \sum_{i = 0}^m Q_i(x)\cdot y_i$ with $Q_i(0)\neq 0$ for some $i \in \{0,1,\dots,m\}$, $x$-degree $D$, $k$ is a natural number and $\gamma \in \F$ is an element of order at least $k$. 
  }
  \KwOut{
    A basis for the affine space of solutions of degree at most $d$ of $Q(x, f(x), \ldots, f(\gamma^m x)) \equiv 0 \mod x^k$. 
  }
  \nonl\hrulefill\\
  \For{$t \in Z_{Q, d}\cup\{-1\}$}{ 
    Run \funcRSolveQ (\autoref{algo:frs-recurse}) on inputs $(Q, d,t)$ to obtain polynomial $f_t(x)$;
  }      
  \KwRet{$f_{-1},f_0, f_1, \ldots, f_{m-1}$.}
  \end{algorithm}
  \end{minipage}

The time complexity of \funcSolveQ (\autoref{algo:frs-solve-original}) is clearly dominated by the time complexity of the $m$ recursive calls to \funcRSolveQ (\autoref{algo:frs-recurse}), which in-turn is a divide and conquer algorithm, where the preprocessing and post-processing cost are both at most $\tilde{O}((D + k)\poly(m))$ field operations, and thus the overall cost is also at most $\tilde{O}((D + k)\poly(m))$.

The correctness of \funcRSolveQ (\autoref{algo:frs-recurse}) follows from the second and the third item of \autoref{lem:frs-divide-conquer} and the correctness of \funcSolveQ (\autoref{algo:frs-solve-original}) follows from the correctness of \funcRSolveQ (\autoref{algo:frs-recurse}) and the first two items of \autoref{lem:frs-structure-of-solution-space-guruswami-wang}. 

We summarise these claims in the following lemma, but skip the formal proof. 

 \begin{lemma}[Time complexity of \funcSolveQ]\label{lem:frs-runtime-correctness}
\funcSolveQ (\autoref{algo:frs-solve-original}) requires at most $\tilde{O}((D+k)\poly(m))$ many field operations in the worst case, and correctly outputs a basis for the solution space of degree $d$ polynomials $f$ satisfying $Q(x, f(x), \ldots, f(\gamma^m x) ) \equiv 0$. 
 \end{lemma}

 Given the results in \autoref{sec:frs-constructing-functional-equations} and \autoref{sec:frs-solving-functional-equations}, it is fairly straightforward to combine them (precisely along the lines of the proof of \autoref{thm:main-large-multiplicity}) to get \autoref{thm:intro-frs-main-large-folding}. We note that while \autoref{thm:prune-krsw} is stated for multiplicity codes, it also holds with the same parameters in a more general sense, and in particular for FRS codes. We skip the details.

\section{Fast decoding up to the Johnson radius}\label{sec:johnson}
In this section, we discuss the main ideas in the proof of \autoref{thm:intro-main-all-multiplicity}. In particular, this yields a nearly-linear-time implementation of Nielsen's algorithm \cite{Nielsen2001} for list decoding univariate multiplicity up to the Johnson radius. The outline of the algorithm is again similar to the polynomial method arguments for decoding Reed-Solomon codes due to Sudan \cite{Sudan1997}.  The main difference between this algorithm and that in the proof of \autoref{thm:intro-main-large-multiplicity} is that we interpolate a bivariate polynomial and not an $(m+2)$-variate polynomial. As a consequence, the error tolerance of the algorithm is low, and we can only  decode upto the Johnson radius. One advantage though is that this works for all multiplicities and not just large enough multiplicities as in \autoref{thm:intro-main-large-multiplicity}. Another technical difference is that we only have to \emph{solve} a bivariate polynomial equation here to recover all the low-degree roots of it. In particular, we do not need to work with differential equations. While the proof is written for univariate multiplicity codes, we remark at the end of the section how the exact same proof works also for folded-RS codes (in fact, any polynomial ideal code).

The main steps of the algorithm involve constructing a bivariate polynomial that \emph{explains} the received word, showing that all close enough codewords satisfy a polynomial equation based on the interpolated bivariate polynomial, and solving these equations to obtain a list of all close enough codewords. We use an algorithm of Roth \& Ruckenstein \cite{RothR2000} and Alekhnovich \cite{Alekhnovich2005} in a blackbox way to solve the equations that appear in this algorithm and our novelty is in constructing the interpolating polynomial in nearly-linear time. The main differences between our algorithm and that of of Alekhnovich \cite{Alekhnovich2005} are in the precise definition of lattices that we work with, and use of Minkoswki's theorem to demonstrate the existence of a short vector in the lattice.   

Now, we describe the details of the interpolation step. 
\subsection{Proof of {\autoref{thm:intro-main-all-multiplicity}}}

The algorithm will proceed broadly as follows. We first find a low-degree ``explanation'' bivariate polynomial $Q$ for the received word. We then show that $Q(x, f(x))$ is identically zero for any polynomial $f$ whose encoding is ``close'' to the received word. Then, the list-decoding problem reduces to finding factors of $Q$ of the form $y-f(x)$. 

Let $R = (\alpha_i, \beta_i(x))_{i = 1}^n$ be a received word\footnote{This algorithm works for all characteristics. For this reason, we work with the ideal-theoretic definition of multiplicity codes given in \cref{fn:mult-ideal-defn}}.  To define our notion of an ``interpolation'' polynomial, first we lay out an interpolating polynomial for the received word as it is.

Let $A(x)$ be the unique univariate polynomial of degree less than $ns$ in $\F[x]$ such that for every $i \in \{1, 2, \ldots, n\}$, 
\begin{equation}\label{eq:interpolation}
A(x) = \beta_i(x) \mod (x-\alpha_i)^s\, .
\end{equation}

The problem of finding $A$ efficiently is nothing but fast Chinese Remainder Algorithm, which has a nearly-linear-time algorithm (essentially the same argument as \autoref{thm:fast-hermite-interpolation}).

\begin{proposition}[{\cite[Algorithm 10.22]{GathenG-MCA}}]
  The interpolating polynomial $A$ can be found in time $(\tilde{O}(ns))$.
\end{proposition}

Let $ r\in \N$ be a positive integer (that we will fix later). Now, we want a $Q$ of $(1, d)$-weighted degree at most some $D$, that satisfies the following.
\begin{equation*}
  (x-\alpha_i)^{s\cdot r} \ \vert \ Q(x, A(x)) \,,\qquad \forall \ i \in \{1, 2, \ldots, n\} \, .
\end{equation*}
 
A natural space to look for such a $Q$ is in the $\F[x]$-linear span of polynomials $(y-A(x))^j\cdot\prod_{i=1}^n (x-\alpha_i)^{s\cdot(r-j)}$ for $j \in \{0,1,\dots,r\}$. For reasons that will become clear later, we will search in a slightly larger space\footnote{The choice of this larger space is inspired by a similar lattice construction due to Coppersmith \cite{Coppersmith1997} in the context of finding small integer solutions to polynomial equations.}. More formally, we consider the following lattice. 

\begin{definition}\label{def:basis-definition}
Let $R = (\alpha_i, \beta_i(x))_{i = 1}^n$ be a received word for a multiplicity code decoder. 

For two positive integers $r\leq u$, let $\mathcal{L}^{(r,u)}_R$ be the $u$-dimensional lattice defined as the $\F[x]$-span of the polynomial vectors $B_0,B_1,\dots, B_u$ defined as follows:
\[ B_j(x,y) := \begin{cases}
            (y-A(x))^j \cdot \prod_{i=1}^n (x-\alpha_i)^{s\cdot(r-j)}\,,& \text{if } 0\leq j \leq r,\\
            (y-A(x))^j\,, & \text{if } r< j \leq u \,.
          \end{cases}
\] 
or equivalently
\[
\mathcal{L}^{(r,u)}_R := \left\{ \sum_{j=0}^r g_j(x)(y-A(x))^{j}\prod_{i=1}^n (x-\alpha_i)^{s\cdot (r-j)} + \sum_{j=r+1}^u g_j(x)(y-A(x))^j\colon g_0, g_1,\dots, g_u \in \F[x] \right\}\,.\qedhere
\]
\end{definition} 
We will eventually be setting $u \approx r \cdot \sqrt{\sfrac{ns}{d}}$, more precisely we set $u$ such that $nsr(r+1)/(u+1) = du$ which would imply that the degree $D$ is bounded above by $\sqrt{nsdr(r+1)}$. We now state the main theorem of this section. 

\begin{theorem}\label{thm:constructing-small-Q-in-ideal}
 Let $R = (\alpha_i, \beta_i(x))_{i = 1}^n$ be a received word for a multiplicity code decoder. Let $A$ be a univariate polynomial of degree at most $ns$ in $\F[x]$ such that for every $i \in \{1, 2, \ldots, n\}$, 
 $A(x) = \beta_i(x) \mod (x-\alpha_i)^s$.
  
 Then, for positive integers $r \leq u$, there exists a non-zero polynomial $Q(x, y)$, with $(1, d)$-weighted degree $D \leq \frac{1}{2}\left(\frac{nsr(r+1)}{u+1}+du\right)$ such that $Q(x,y) \in \mathcal{L}_R^{(r,u)}$.

 Moreover, there is a deterministic algorithm that when given $R$ as input, runs in time $\tilde{O}(ns \cdot \poly(u,ns/d))$ and outputs such a polynomial $Q$ as a list of coefficients. 
\end{theorem}

The proof proceeds along the lines of Alekhnovich's proof. We first use Minkowski's theorem \autoref{thm:minkowski} to show that there exists a non-zero polynomial with small $(1,d)$-weighted degree and then use Alekhnovich's algorithm (\autoref{thm:alekhnovich shortest vector}) to compute this polynomial. 

The following proposition shows the existence of a non-zero polynomial of low $(1,d)$-weighted degree in the lattice.
\begin{proposition}\label{prop:small-Q-in-ideal}
  Let $\mathcal{L}_R^{(r,u)}$ be the lattice constructed as in \autoref{def:basis-definition} for some received word $R$ and parameters $r \leq u$. Then there exists a non-zero polynomial $Q$ in the lattice $\mathcal{L}_R^{(r,u)}$ with $(1,d)$-weighted degree at most $\frac{1}{2}\left(\frac{nsr(r+1)}{u+1}+du\right)$.
\end{proposition}
\begin{proof}
It will be convenient to work with the $(u+1)$-dimensional lattice $\widehat{\mathcal{L}}_R^{(r,u)}$ spanned by the following $(u+1)$ polynomial vectors.
\begin{equation}\label{eq:lhatbasis} 
  \hat{B}_j(x,y) := \begin{cases}
  (y\cdot x^d-A(x))^j \cdot \prod_{i=1}^n (x-\alpha_i)^{s\cdot(r-j)}\,,& \text{if } 0\leq j \leq r,\\
  (y\cdot x^d-A(x))^j\,, & \text{if } r< j \leq u \,.
\end{cases}
\end{equation} 
There is a natural 1-1 correspondence between the polynomials in the two lattices $\mathcal{L}_R^{(r,u)}$ and $\widehat{\mathcal{L}}_R^{(r,u)}$ given by $y \longleftrightarrow y\cdot x^d$. Clearly, the non-zero polynomial in $\widehat{\mathcal{L}}_R^{(r,u)}$ with smallest $x$-degree corresponds to the non-zero polynomial in $\mathcal{L}_R^{(r,u)}$ with smallest $(1,d)$-weighted degree. We now show that there exists a non-zero polynomial $\hat{Q} \in \widehat{\mathcal{L}}_R^{(r,u)}$ with $x$-degree at most $\frac{1}{2}\left(\frac{nsr(r+1)}{u+1}+du\right)$ which by the above correspondence implies that there is a non-zero polynomial $Q$ in the lattice $\mathcal{L}_R^{(r,u)}$ with $(1,d)$-weighted degree at most $\frac{1}{2}\left(\frac{nsr(r+1)}{u+1}+du\right)$. We note that while we have been viewing polynomials in the lattice as bivariate polynomials, we can equivalently view any such polynomial $\hat{Q}$ as an $(u + 1)$-dimensional column vector with entries in the ring $\F[x]$, where the coordinates are labelled $(0, 1, \ldots, u)$ and for $j \leq u$, the $j^{th}$ coordinate corresponds to the coefficient of $y^j$ in the polynomial $\hat{Q}$. For instance, if $\hat{Q}(x,y) = \sum_{j=0}^u q_j(x) \cdot y^j$, then this corresponds to the transpose of $(q_0, q_1, \ldots, q_u)$. Moreover, we note that the vectors corresponding to the generators of the lattice $\widehat{{\cal L}}_R^{(r,u)}$, namely the polynomials $(y\cdot x^d-A(x))^j \cdot \prod_{i=1}^n (x-\alpha_i)^{s\cdot(r-j)}$ for $0\leq j \leq r$ and $(y\cdot x^d-A(x))^j$ for $r< j \leq u$ form a lower triangular matrix with non-zero diagonal entries, and hence are linearly independent over $\F(x)$. In other words, the lattice $\widehat{{\cal L}}_R^{(r,u)}$ is full rank. We also know the diagonal entries precisely, the $x$-degree of the $j^{th}$ diagonal entry is exactly ${ns(r-j)}+dj$ for $0 \leq j\leq r$ and ${{dj}}$ for $r<j\leq u$. Hence, the determinant of this matrix is a polynomial of degree at most (in fact, it is equal to) $nsr(r+1)/2 + du(u+1)/2$. From Minkowski's Theorem (\autoref{thm:minkowski}), we have that there is a non-zero vector $\hat{Q}(x,y)$ in the $(u+1)$-dimensional lattice $\widehat{\cal L}_R^{(r,u)}$ with every entry of degree at most $\sfrac{1}{2}\left(\frac{nsr(r+1)}{u+1} + du\right)$. The corresponding non-zero polynomial $Q(x,y)$ in the lattice $\mathcal{L}_R^{(r,u)}$ then has $(1,d)$-weighted degree at most $\sfrac{1}{2}\left(\frac{nsr(r+1)}{u+1} + du\right)$.
\end{proof}

We are now ready to prove the main theorem of the section.

\begin{proof}[Proof of \autoref{thm:constructing-small-Q-in-ideal}]
  The existence of a non-zero $Q$ with small $(1,d)$-weighted degree follows from \autoref{prop:small-Q-in-ideal}. We can then invoke Alekhnovich's algorithm (\autoref{thm:alekhnovich shortest vector}) on the set of generators $\hat{B_0}, \hat{B_1},\ldots, \hat{B_u}$ for the lattice $\widehat{\mathcal{L}}_R^{(r,u)}$ (as given by \eqref{eq:lhatbasis}) to construct the non-zero $\hat{Q}$ of least $x$-degree in time $\tilde{O}(nsr \cdot \poly(u))$. Here, we use the fact that the lattice is given by $(u+1)$ vectors in dimensions $(u+1$) over the ring $\F[x]$ and each entry of the vectors is a polynomial of degree at  most $nsr$. The corresponding non-zero polynomial  $Q$ in the lattice $\mathcal{L}_R^{(r,u)}$ has the least $(1,d)$-weighted degree which by the first part of the theorem is at most  $\sfrac{1}{2}\left(\frac{nsr(r+1)}{u+1} + du\right)$.
\end{proof}

It remains to be shown that codewords $f$ close to the received word cause $Q(x, f(x))$ to be identically zero and we do this in the next lemma. 

\begin{lemma}\label{lem:close-codewords}
  Let $R = (\alpha_i, \beta_i(x))_{i = 1}^n$ be a received word for a multiplicity code decoder, and let $Q(x, y) \in \mathcal{L}^{(r,u)}_R$ be a non-zero polynomial with $(1, d)$-weighted degree at most $D$ for some positive integers $r\leq u$.  
  
  Then, for every polynomial $f \in \F[x]$ of degree at most $d$ such that $\enc_{s, \vec{\alpha}}(f)$ agrees with $R$ on more than $\sfrac{D}{s\cdot r}$ values of $i$, $Q(x, f(x))$ is identically zero. 

  \end{lemma}
  \begin{proof}
  Define $P(x) := Q(x, f(x))$. Since $Q(x, y)$ has $(1, d)$-weighted degree at most $D$ and $f$ has degree $d$, $P(x)$ must have $x$-degree at most $D$. 

 We first show that for every point $(\alpha_i,\beta_i(x))$ of agreement between the received word $R$ and $\enc_{s, \vec{\alpha}}(f)$, we have $(x-\alpha_i)^{s\cdot r}$ divides $Q(x, f(x))$. To see this, first notice that $A(x) = \beta_i(x) \mod (x-\alpha_i)^2$ as well as $f(x) = \beta_i(x) \mod (x-\alpha_i)^s$ and hence, $(x-\alpha_i)^s$ divides $(f(x)-A(x))$. Hence, $(x-\alpha_i)^{s(j+j')}$ divides $(f(x)-A(x))^j \cdot \prod_{i=1}^n (x-\alpha_i)^{s\cdot j'}$ for any pair of non-negative integers $j$ and $j'$. Hence, $(x-\alpha_i)^{sr}$ divides $B_j(x,f(x))$ for each basis vector $B_j$ of the lattice $\mathcal{L}_R^{(r,u)}$. Since $Q(x,f(x)) = \sum g_j(x)\cdot B_j(x,f(x))$ we have that $(x-\alpha_i)^{sr}$ divides $Q(x,f(x))$. 
 
 Thus, if $R$ and $\enc_{s, \vec{\alpha}}(f)$ have at least $t > \sfrac{D}{rs}$ agreement points, then $P$ has a total of $t\cdot rs > D$ zeros (counting multiplicities). However, from earlier, $P$ has degree at most $D$, so it must be identically zero.
\end{proof}

Then, the final step is to find factors of $Q(x, y)$, which are of the form $(y-f(x))$. For this, we invoke an algorithm of Roth \& Ruckenstein \cite{RothR2000} and Alekhnovich \cite{Alekhnovich2005}. 
\begin{theorem}[{\cite{RothR2000}, \cite[Theorem 1.2]{Alekhnovich2005}}]\label{thm:factorization}
There is a deterministic algorithm that given a bivariate polynomial $Q(x,y)$ finds all polynomials $f(x)$ such that $Q(x,f(x)) \equiv 0$ in time $\tilde{O}(D\cdot\poly(\ell))$, where $D$ is the $x$-degree of $Q$ and $\ell$ is the $y$-degree of $Q$. 
\end{theorem}

The proof of \autoref{thm:intro-main-all-multiplicity} now immediately follows from a combination of \autoref{thm:constructing-small-Q-in-ideal}, \autoref{lem:close-codewords} and \autoref{thm:factorization} by setting $r,u$ such that $u(u+1)=r(r+1)ns/d$ and $r=O(\sfrac{1}{\epsilon})$. We skip the details. 

This proof verbatim extends for any \emph{polynomial ideal code} \cite{BhandariHKS2024-affineFRS}.  A polynomial ideal code is specified by a family of $n$ univariate monic polynomials $E_1, E_2,\ldots, E_n$ of degree exactly $s$ and the $n$-symbol encoding of a polynomial $f \in \F_{\leq d}[x]$ is specified by $\left(f(x) \mod E_i(x)\right)_{i=1}^n$. A polynomial ideal code is thus a common generalization of Reed-Solomon codes ($E_i(x) = (x-\alpha_i)$), folded-RS codes ($E_i(x)= \prod_{j=0}^{s-1}(x-\gamma^j \alpha_i)$) and univariate multiplicity codes ($E_i(x)= (x-\alpha_i)^s$). The only difference in the proof is replacing \eqref{eq:interpolation} by $A(x) = \beta_i(x) \mod E_i(x)$. 

\section*{Acknowledgements}
Some of the discussions of the second and the third authors leading up to this work happened while they were visiting the Homi Bhabha Center for Science Education (HBCSE), Mumbai. We are thankful to Arnab Bhattacharya and the HBCSE staff for their warm and generous hospitality and for providing an inviting and conducive atmosphere for these discussions.

We are also thankful to Madhu Sudan for comments and suggestions on an earlier draft of the paper that helped improve the writing. We thank Noga Ron-Zewi for bringing to our attention the work on nearly-linear-time list-decoding algorithms for certain families of tensor-product codes \cite{HemenwayRW2020,KoppartyRRSS2021}. We thank  Swastik Kopparty for many helpful discussions on the theme of list decoding over the years, and in particular, for patiently answering our many questions related to the results in \cite{HemenwayRW2020,KoppartyRRSS2021}. 

Finally, we are thankful to the anonymous reviewers at STOC 2024, whose comments were helpful in cleaning up parts of the presentation of the paper. 

\addcontentsline{toc}{section}{References}
{\small 
  \bibliographystyle{prahladhurl}
  \bibliography{fastdecoding-bib}

\begin{thebibliography}{KRRSS21}

\bibitem[AEL95]{AlonEL1995}
\textsc{Noga Alon}, \textsc{Jeff Edmonds}, and \textsc{Michael Luby}.
\newblock \href{https://doi.org/10.1109/SFCS.1995.492581} {\emph{Linear time
  erasure codes with nearly optimal recovery}}.
\newblock In \textsc{Prabhakar Raghavan}, ed., \emph{Proc.\ $36$th IEEE Symp.\
  on Foundations of Comp.\ Science (FOCS)}, pages 512--519. 1995.

\bibitem[AGL23]{AlrabiahGL2023}
\textsc{Omar Alrabiah}, \textsc{Venkatesan Guruswami}, and \textsc{Ray Li}.
\newblock \emph{Randomly punctured {R}eed-{S}olomon codes achieve list-decoding
  capacity over linear-sized fields}, 2023.
\newblock (manuscript).
\newblock \href{http://arxiv.org/abs/2304.09445}{\path{arXiv:2304.09445}},
  \href{https://eccc.weizmann.ac.il/eccc-reports/2023/TR23-125}{\path{eccc:2023/TR23-125}}.

\bibitem[AK91]{AbramovK1991}
\textsc{Sergei~A. Abramov} and \textsc{K.~Yu. Kvashenko}.
\newblock \href{https://doi.org/10.1145/120694.120735} {\emph{Fast algorithms
  to search for the rational solutions of linear differential equations with
  polynomial coefficients}}.
\newblock In \textsc{Stephen~M. Watt}, ed., \emph{Proc.\ International Symp.\
  Symbolic and Algebraic Computation (ISSAC)}, pages 267--270. {ACM}, 1991.

\bibitem[AL96]{AlonL1996}
\textsc{Noga Alon} and \textsc{Michael Luby}.
\newblock \href{https://doi.org/10.1109/18.556669} {\emph{A linear time
  erasure-resilient code with nearly optimal recovery}}.
\newblock IEEE Trans.\ Inform.\ Theory, 42(6):1732--1736, 1996.
\newblock (Preliminary version in \emph{36th FOCS}, 1995).

\bibitem[Ale05]{Alekhnovich2005}
\textsc{Michael Alekhnovich}.
\newblock \href{https://doi.org/10.1109/TIT.2005.850097} {\emph{Linear
  {D}iophantine equations over polynomials and soft decoding of reed-solomon
  codes}}.
\newblock IEEE Trans.\ Inform.\ Theory, 51(7):2257--2265, 2005.
\newblock (Preliminary version in \emph{43rd FOCS}, 2002).

\bibitem[BCS05]{BostanCS2005}
\textsc{Alin Bostan}, \textsc{Thomas Cluzeau}, and \textsc{Bruno Salvy}.
\newblock \href{https://doi.org/10.1145/1073884.1073893} {\emph{Fast algorithms
  for polynomial solutions of linear differential equations}}.
\newblock In \textsc{Manuel Kauers}, ed., \emph{Proc.\ International Symp.\
  Symbolic and Algebraic Computation (ISSAC)}, pages 45--52. {ACM}, 2005.

\bibitem[Ber67]{Berlekamp1967}
\textsc{Elwyn~R. Berlekamp}.
\newblock \href{https://doi.org/10.1109/TIT.1968.1054109} {\emph{Nonbinary
  {BCH} decoding}}.
\newblock In \emph{Proc. IEEE International Symposium on Information Theory
  (ISIT)}. 1967.

\bibitem[BHKS24a]{BhandariHKS2024-mgrid}
\textsc{Siddharth Bhandari}, \textsc{Prahladh Harsha}, \textsc{Mrinal Kumar},
  and \textsc{Madhu Sudan}.
\newblock \href{https://doi.org/10.1109/TIT.2023.3306849} {\emph{Decoding
  multivariate multiplicity codes over product sets}}.
\newblock IEEE Trans.\ Inform.\ Theory, 70(1):154--169, 2024.
\newblock (Preliminary version in \emph{53rd STOC}, 2021).
\newblock \href{http://arxiv.org/abs/2012.01530}{\path{arXiv:2012.01530}},
  \href{https://eccc.weizmann.ac.il/eccc-reports/2020/TR20-179}{\path{eccc:2020/TR20-179}}.

\bibitem[BHKS24b]{BhandariHKS2024-affineFRS}
---{}---{}---.
\newblock \href{https://doi.org/10.1109/TIT.2023.3345890}
  {\emph{Ideal-theoretic explanation of capacity-achieving decoding}}.
\newblock IEEE Trans.\ Inform.\ Theory, 70(2):1107--1123, 2024.
\newblock (Preliminary version in \emph{25th RANDOM}, 2021).
\newblock \href{http://arxiv.org/abs/2103.07930}{\path{arXiv:2103.07930}},
  \href{https://eccc.weizmann.ac.il/eccc-reports/2021/TR21-036}{\path{eccc:2021/TR21-036}}.

\bibitem[Cop97]{Coppersmith1997}
\textsc{Don Coppersmith}.
\newblock \href{https://doi.org/10.1007/S001459900030} {\emph{Small solutions
  to polynomial equations, and low exponent {RSA} vulnerabilities}}.
\newblock J. Cryptol., 10(4):233--260, 1997.
\newblock (Preliminary version in \emph{EUROCRYPT}, 1996).

\bibitem[GG13]{GathenG-MCA}
\textsc{Joachim von~zur Gathen} and \textsc{J{\"{u}}rgen Gerhard}.
\newblock \href{https://doi.org/10.1017/CBO9781139856065} {\emph{Modern
  Computer Algebra}}.
\newblock Cambridge University Press, 3 edition, 2013.

\bibitem[GI03]{GuruswamiI2003}
\textsc{Venkatesan Guruswami} and \textsc{Piotr Indyk}.
\newblock \href{https://doi.org/10.1145/780542.780562} {\emph{Linear time
  encodable and list decodable codes}}.
\newblock In \textsc{Lawrence~L. Larmore} and \textsc{Michel~X. Goemans}, eds.,
  \emph{Proc.\ $35$th ACM Symp.\ on Theory of Computing (STOC)}, pages
  126--135. 2003.

\bibitem[GR08]{GuruswamiR2008}
\textsc{Venkatesan Guruswami} and \textsc{Atri Rudra}.
\newblock \href{https://doi.org/10.1109/TIT.2007.911222} {\emph{Explicit codes
  achieving list decoding capacity: Error-correction with optimal redundancy}}.
\newblock IEEE Trans.\ Inform.\ Theory, 54(1):135--150, 2008.
\newblock (Preliminary version in \emph{38th STOC}, 2006).
\newblock \href{http://arxiv.org/abs/cs/0511072}{\path{arXiv:cs/0511072}},
  \href{https://eccc.weizmann.ac.il/eccc-reports/2005/TR05-133}{\path{eccc:2005/TR05-133}}.

\bibitem[GS99]{GuruswamiS1999}
\textsc{Venkatesan Guruswami} and \textsc{Madhu Sudan}.
\newblock \href{https://doi.org/10.1109/18.782097} {\emph{Improved decoding of
  {R}eed-{S}olomon and algebraic-geometry codes}}.
\newblock IEEE Trans.\ Inform.\ Theory, 45(6):1757--1767, 1999.
\newblock (Preliminary version in \emph{39th FOCS}, 1998).
\newblock
  \href{https://eccc.weizmann.ac.il/eccc-reports/1998/TR98-043}{\path{eccc:1998/TR98-043}}.

\bibitem[GW13]{GuruswamiW2013}
\textsc{Venkatesan Guruswami} and \textsc{Carol Wang}.
\newblock \href{https://doi.org/10.1109/TIT.2013.2246813}
  {\emph{Linear-algebraic list decoding for variants of {R}eed-{S}olomon
  codes}}.
\newblock IEEE Trans.\ Inform.\ Theory, 59(6):3257--3268, 2013.
\newblock (Preliminary version in \emph{26th {IEEE} Conference on Computational
  Complexity}, 2011 and \emph{15th RANDOM}, 2011).
\newblock
  \href{https://eccc.weizmann.ac.il/eccc-reports/2012/TR12-073}{\path{eccc:2012/TR12-073}}.

\bibitem[HRW20]{HemenwayRW2020}
\textsc{Brett Hemenway}, \textsc{Noga Ron{-}Zewi}, and \textsc{Mary Wootters}.
\newblock \href{https://doi.org/10.1137/17M116149X} {\emph{Local list recovery
  of high-rate tensor codes and applications}}.
\newblock SIAM J. Comput., 49(4), 2020.
\newblock (Preliminary version in \emph{58th FOCS}, 2017).
\newblock \href{http://arxiv.org/abs/1706.03383}{\path{arXiv:1706.03383}},
  \href{https://eccc.weizmann.ac.il/eccc-reports/2017/TR17-104}{\path{eccc:2017/TR17-104}}.

\bibitem[Kop14]{Kopparty2014}
\textsc{Swastik Kopparty}.
\newblock \href{https://doi.org/10.1090/conm/625/12497} {\emph{Some remarks on
  multiplicity codes}}.
\newblock In \textsc{Alexander Barg} and \textsc{Oleg~R. Musin}, eds.,
  \emph{Discrete Geometry and Algebraic Combinatorics}, volume 625 of
  \emph{Contemporary Mathematics}, pages 155--176. AMS, 2014.
\newblock \href{http://arxiv.org/abs/1505.07547}{\path{arXiv:1505.07547}}.

\bibitem[Kop15]{Kopparty2015}
---{}---{}---.
\newblock \href{https://doi.org/10.4086/toc.2015.v011a005} {\emph{List-decoding
  multiplicity codes}}.
\newblock Theory of Computing, 11:149--182, 2015.
\newblock
  \href{https://eccc.weizmann.ac.il/eccc-reports/2012/TR12-044}{\path{eccc:2012/TR12-044}}.

\bibitem[Kra03]{Krachkovsky2003}
\textsc{Victor~Yu. Krachkovsky}.
\newblock \href{https://doi.org/10.1109/TIT.2003.819333}
  {\emph{{R}eed-{S}olomon codes for correcting phased error bursts}}.
\newblock IEEE Trans.\ Inform.\ Theory, 49(11):2975--2984, 2003.

\bibitem[KRRSS21]{KoppartyRRSS2021}
\textsc{Swastik Kopparty}, \textsc{Nicolas Resch}, \textsc{Noga Ron{-}Zewi},
  \textsc{Shubhangi Saraf}, and \textsc{Shashwat Silas}.
\newblock \href{https://doi.org/10.1109/TIT.2020.3023962} {\emph{On list
  recovery of high-rate tensor codes}}.
\newblock IEEE Trans.\ Inform.\ Theory, 67(1):296--316, 2021.
\newblock (Preliminary version in \emph{23rd RANDOM}, 2019).
\newblock
  \href{https://eccc.weizmann.ac.il/eccc-reports/2019/TR19-080}{\path{eccc:2019/TR19-080}}.

\bibitem[KRSW18]{KoppartyRSW2018}
\textsc{Swastik Kopparty}, \textsc{Noga Ron{-}Zewi}, \textsc{Shubhangi Saraf},
  and \textsc{Mary Wootters}.
\newblock \href{https://doi.org/10.1109/FOCS.2018.00029} {\emph{Improved
  decoding of {F}olded {R}eed-{S}olomon and {M}ultiplicity codes}}.
\newblock In \textsc{Mikkel Thorup}, ed., \emph{Proc.\ $59$th IEEE Symp.\ on
  Foundations of Comp.\ Science (FOCS)}, pages 212--223. 2018.
\newblock \href{http://arxiv.org/abs/1805.01498}{\path{arXiv:1805.01498}},
  \href{https://eccc.weizmann.ac.il/eccc-reports/2018/TR18-091}{\path{eccc:2018/TR18-091}}.

\bibitem[KRSW23]{KoppartyRSW2023}
---{}---{}---.
\newblock \href{https://doi.org/10.1137/20M1370215} {\emph{Improved list
  decoding of {F}olded {R}eed-{S}olomon and {M}ultiplicity codes}}.
\newblock SIAM J. Comput., 52(3):794--840, 2023.
\newblock (Preliminary version in \emph{59th FOCS}, 2018).
\newblock \href{http://arxiv.org/abs/1805.01498}{\path{arXiv:1805.01498}},
  \href{https://eccc.weizmann.ac.il/eccc-reports/2018/TR18-091}{\path{eccc:2018/TR18-091}}.

\bibitem[KSY14]{KoppartySY2014}
\textsc{Swastik Kopparty}, \textsc{Shubhangi Saraf}, and \textsc{Sergey
  Yekhanin}.
\newblock \href{https://doi.org/10.1145/2629416} {\emph{High-rate codes with
  sublinear-time decoding}}.
\newblock J. ACM, 61(5):28:1--28:20, 2014.
\newblock (Preliminary version in \emph{43rd STOC}, 2011).
\newblock
  \href{https://eccc.weizmann.ac.il/eccc-reports/2010/TR10-148}{\path{eccc:2010/TR10-148}}.

\bibitem[Kun74]{Kung1974}
\textsc{Hsiang-Tsung Kung}.
\newblock \href{https://doi.org/10.1007/BF01436917} {\emph{On computing
  reciprocals of power series}}.
\newblock Numer.\ Math., 22(5):341--348, 1974.

\bibitem[Mas69]{Massey1969}
\textsc{James~L. Massey}.
\newblock \href{https://doi.org/10.1109/TIT.1969.1054260} {\emph{Shift-register
  synthesis and {BCH} decoding}}.
\newblock IEEE Trans.\ Inform.\ Theory, 15(1):122--127, 1969.

\bibitem[Nie01]{Nielsen2001}
\textsc{Rasmus~Refslund Nielsen}.
\newblock \href{https://orbit.dtu.dk/en/publications/list-decoding-of-
  linear-block-codes} {\emph{List decoding of linear block codes}}.
\newblock Ph.D. thesis, Technical University of Denmark, 2001.

\bibitem[PV05]{ParvareshV2005}
\textsc{Farzad Parvaresh} and \textsc{Alexander Vardy}.
\newblock \href{https://doi.org/10.1109/SFCS.2005.29} {\emph{Correcting errors
  beyond the {G}uruswami-{S}udan radius in polynomial time}}.
\newblock In \textsc{\'{E}va Tardos}, ed., \emph{Proc.\ $46$th IEEE Symp.\ on
  Foundations of Comp.\ Science (FOCS)}, pages 285--294. 2005.

\bibitem[RR00]{RothR2000}
\textsc{Ron~M. Roth} and \textsc{Gitit Ruckenstein}.
\newblock \href{https://doi.org/10.1109/18.817522} {\emph{Efficient decoding of
  {R}eed-{S}olomon codes beyond half the minimum distance}}.
\newblock IEEE Trans.\ Inform.\ Theory, 46(1):246--257, 2000.

\bibitem[RSTW78]{ReedSTW1978}
\textsc{Irving~S. Reed}, \textsc{Robert~A. Scholtz}, \textsc{Trieu{-}Kien
  Truong}, and \textsc{Lloyd~R. Welch}.
\newblock \href{https://doi.org/10.1109/TIT.1978.1055816} {\emph{The fast
  decoding of {R}eed-{S}olomon codes using {F}ermat theoretic transforms and
  continued fractions}}.
\newblock IEEE Trans.\ Inform.\ Theory, 24(1):100--106, 1978.

\bibitem[RT97]{RosenbloomT1997}
\textsc{M.~Yu Rosenbloom} and \textsc{Michael~Anatol\'{e}vich Tsfasman}.
\newblock \href{http://mi.mathnet.ru/ppi359}
  {\emph{\foreignlanguage{russian}{Коды для m-метрики}
  ({R}ussian) [{C}odes for the $m$-metric]}}.
\newblock Probl.\ Peredachi Inf., 33(1):55--63, 1997.
\newblock (English translation in \emph{Problems Inform. Transmission},
  33(1):45--52, 1997).

\bibitem[Sie72]{Sieveking1972}
\textsc{Malte Sieveking}.
\newblock \href{https://doi.org/10.1007/BF02242389} {\emph{An algorithm for
  division of powerseries}}.
\newblock Computing, 10(1-2):153--156, 1972.

\bibitem[Sin91]{Singer1991}
\textsc{Michael~F. Singer}.
\newblock \href{https://doi.org/10.1016/S0747-7171(08)80048-X}
  {\emph{{L}iouvillian solutions of linear differential equations with
  {L}iouvillian coefficients}}.
\newblock J. Symb. Comput., 11(3):251--273, 1991.
\newblock (Preliminary version in \emph{3rd Computers and Mathematics} 1989).

\bibitem[SS96]{SipserS1996}
\textsc{Michael Sipser} and \textsc{Daniel~A. Spielman}.
\newblock \href{https://doi.org/10.1109/18.556667} {\emph{Expander codes}}.
\newblock IEEE Trans.\ Inform.\ Theory, 42(6):1710--1722, November 1996.
\newblock (Preliminary version in \emph{35th FOCS}, 1994).

\bibitem[ST23]{ShangguanT2023}
\textsc{Chong Shangguan} and \textsc{Itzhak Tamo}.
\newblock \href{https://doi.org/10.1137/20M138795X} {\emph{Generalized
  {S}ingleton bound and list-decoding {R}eed-{S}olomon codes beyond the
  {J}ohnson radius}}.
\newblock SIAM J. Comput., 52(3):684--717, 2023.
\newblock (Preliminary version in \emph{52nd STOC}, 2020).
\newblock \href{http://arxiv.org/abs/1911.01502}{\path{arXiv:1911.01502}}.

\bibitem[Sud97]{Sudan1997}
\textsc{Madhu Sudan}.
\newblock \href{https://doi.org/10.1006/jcom.1997.0439} {\emph{Decoding of
  {R}eed-{S}olomon codes beyond the error-correction bound}}.
\newblock J. Complexity, 13(1):180--193, 1997.
\newblock (Preliminary version in \emph{37th FOCS}, 1996).

\bibitem[Tam23]{Tamo2024}
\textsc{Itzhak Tamo}.
\newblock \emph{Tighter list-size bounds for list-decoding and recovery of
  folded {R}eed-{S}olomon and multiplicity codes}, 2023.
\newblock (manuscript).
\newblock \href{http://arxiv.org/abs/2312.17097}{\path{arXiv:2312.17097}}.

\end{thebibliography}
}


\end{document}